\documentclass[11pt,aps,superscriptaddress,floatfix]{revtex4}
\usepackage{graphicx}
\usepackage{epsfig}
\usepackage{amsmath,amssymb}
\usepackage{color}

\newcommand{\be}{\begin{eqnarray}}
\newcommand{\ee}{\end{eqnarray}}
\newcommand\del{\partial}
\def\conj#1{{{#1}^{*}}}

\newcommand{\el}{\nonumber \hfill \\}
\newcommand{\mat}{\left ( \begin{array}{cc}}
\newcommand{\emat}{\end{array} \right )}
\newcommand{\matf}{\left ( \begin{array}{cccc}}
\newcommand{\ematf}{\end{array} \right )}            
\newcommand{\matt}{\left \begin{array}{ccc}}
\newcommand{\ematt}{\end{array} \right )}
\newcommand{\vect}{\left ( \begin{array}{c}}
\newcommand{\evect}{\end{array} \right )}
\newcommand{\Tr}{{\rm Tr}}

\newcommand{\nn}{\nonumber }

\newcommand{\tr}{{\rm Tr}}
\def\dslash{\makebox[0cm][l]{$\,/$}D}

\newcommand{\e}{{\mathrm e}}
\newcommand{\cD}{{\cal D}}
\newcommand{\cZ}{{\cal Z}}
\newcommand{\cN}{{\cal N}}
\newcommand{\cH}{{\cal H}}
\newcommand{\cK}{{\cal K}}
\newcommand{\zs}{\conj{z}}
\newcommand{\zssq}{\conjsq{z}}
\newcommand{\hz}{\hat{z}}

\newcommand{\hm}{\hat{m}}
\newcommand{\hmu}{\hat{\mu}}
\def\conj#1{{{#1}^{*}}}
\def\conjsq#1{{{#1}^{* 2}}}

\definecolor{Bittersweet}   {cmyk}{0,0.75,1,0.24}

\DeclareMathOperator{\sgn}{sgn}

\begin{document}

\title{ Chiral Condensate at Nonzero Chemical Potential in the Microscopic Limit of QCD}

\author{J.C. Osborn}
\affiliation{Argonne Leadership Computing Facility, 9700 S. Cass Avenue,
Argonne, IL 60439, USA}
\affiliation{Center for Computational Science, Boston University,
Boston, MA 02215, USA}

\author{K. Splittorff}
\affiliation{The Niels Bohr Institute, Blegdamsvej 17, DK-2100, Copenhagen {\O}, Denmark}

\author{J.J.M. Verbaarschot}
\affiliation{Department of Physics and Astronomy, SUNY, Stony Brook,
 New York 11794, USA}

\date{\today}
\begin{abstract}
  The chiral condensate in QCD at zero temperature does not depend on
  the quark chemical potential (up to one third the nucleon mass),
  whereas the spectral density of the Dirac operator shows a strong
  dependence on the chemical potential.  The cancellations which make
  this possible also occur on the microscopic scale, where they can be
  investigated by means of a random matrix model. We show that they
  can be understood in terms of orthogonality properties of orthogonal
  polynomials. In the strong non-Hermiticity limit they are related to
  integrability properties of the spectral density.  As a by-product
  we find exact analytical expressions for the partially quenched
  chiral condensate in the microscopic domain at nonzero chemical
  potential.
\end{abstract}
\maketitle

\section{Introduction}

One of the simplest questions one can ask regarding the phase diagram of QCD
is: What is the dependence of the chiral condensate on the quark chemical
potential at zero temperature?  This simple question has an equally simple
answer: At low temperature, the vacuum of QCD is dominated by the pions 
and since they have zero baryon charge, the free energy remains
$\mu$-independent until $\mu$ surpasses a third of the nucleon mass.  
Despite its simplicity, even this answer 
is exceedingly hard to verify by direct lattice QCD computations. To see
why, let us consider the relationship between (the magnitude of)
the chiral condensate 
\be
\label{SigmaFromZ}
\Sigma(m) \equiv \frac 1V \frac{d}{dm}\log Z(m) = \frac 1V
\left\langle \Tr \frac{1}{\dslash+\mu\gamma_0+m}\right\rangle
\ee
and the spectral density $\rho$ of the Euclidean 
Dirac operator, $\dslash+\mu\gamma_0$. 
At nonzero chemical potential, the Dirac operator is non-Hermitian and  
the spectral density spreads out from the
imaginary axis. Since the chiral condensate is related to 
this $\mu$-dependent spectral density by
\be
\label{rhoSigma} 
\Sigma(m) = \frac 1V
\int_\mathbb{C} d^2z \, \frac{1}{z+m} \, \rho(z,z^*,m;\mu)
\ee
it is far from obvious that the chiral condensate  obtained this way 
is independent of $\mu$ (for $\mu<m_N/3$ and zero temperature).

The quark mass enters the support of the Dirac spectrum when
$\mu \ge m_\pi/2$ \cite{Gibbs,TV}, and by an electrostatic analogy
\cite{Barbour} one could conclude that the chiral condensate becomes
$\mu$-dependent and always vanishes in the chiral limit $(m\to0)$ when the
chemical potential is non-zero. While this conclusion is correct for
quenched QCD, it fails completely in the unquenched case where, as
argued above, the unquenched chiral condensate is independent of $\mu$.

In the unquenched case, the generally accepted picture for two decades 
was that chiral symmetry breaking requires that the Dirac spectrum should  
accumulate to a non-zero density on the imaginary axis (like the Banks-Casher 
relation \cite{BC} at $\mu=0$) despite the fact that
the anti-Hermiticity of the Dirac operator is explicitly broken by the
chemical potential (see \cite{Barbour} for a discussion of this point). 
Indeed, intricate cancellations due to dynamical fermions set
in when $\mu$ becomes greater than $m_\pi/2$. The eigenvalues, however, do
not accumulate on the imaginary axis.
Direct input \cite{O,AOSV} from 
the microscopic domain of QCD shows that  the discontinuity in the
chiral condensate arises due to strongly oscillating terms in the eigenvalue
density \cite{OSV}. The same mechanism is also responsible for chiral
symmetry breaking in one dimensional QCD at nonzero $\mu$ \cite{RV}. 
Because of the oscillations, the unquenched eigenvalue density is not real and
positive. This fact is a direct a consequence of the sign problem: since the
measure of the Euclidean QCD partition function includes a complex valued
fermion determinant the expectation value of a real and positive function,
such as $\sum_k\delta^{2}(z-z_k)$, needs not be real and
positive.  

The microscopic domain \cite{SV} is the region where 
the  eigenvalues of the Dirac operator are in the domain
\be
| z| \ll  \frac 1{\Lambda_{\rm QCD}\sqrt V} ,
\ee
while the volume is much larger than $\Lambda_{\rm QCD}^{-4}$. The quark
masses may or may not be in this region. In this region the QCD partition
function can be expressed in terms of microscopic scaling variables
\be
\hat z = z V \Sigma, ~~ \hat m = m V \Sigma ~~ \mathrm{and} ~~
 \hat \mu  =\mu  F_\pi \sqrt V, 
\label{micros}
\ee
which stay fixed in the thermodynamic limit. This scaling limit is known as
the microscopic limit. The oscillations
in the spectral density with a period proportional to the inverse volume
are resolved at this scale.
The cancellations of these oscillating contributions with an amplitude
that grows exponentially large with the volume results
in a $\mu$-independent chiral
condensate. They give us a direct insight into problems faced by
lattice QCD at nonzero chemical potential.

In \cite{OSV} the $\mu$-independence of the 
chiral condensate was established using an asymptotic approximation to the
exact expression for the microscopic spectral density. The argument was
carried through by complex contour integrations. In the present paper
we extend this result to the complete 
mass dependence of the chiral condensate in the microscopic domain.

\begin{figure}[t]
  \label{fig:rhoPD}
  \begin{center}
    \epsfig{height=8cm,file=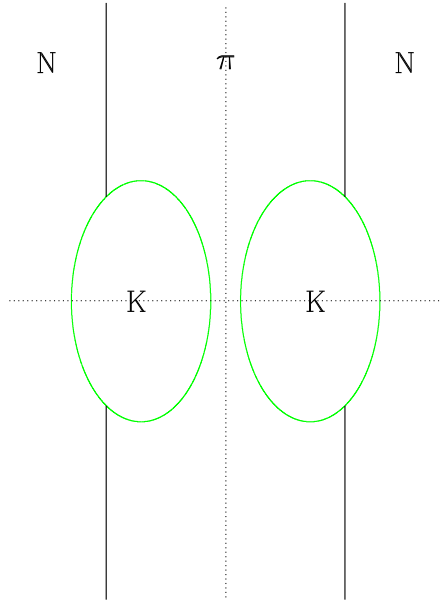}
    \caption{
      Phase diagram of the Dirac spectrum in the complex $z$
      plane for one dynamical flavor. The support of the quenched
      spectrum is between the two vertical black lines.  Unquenching
      introduces the oscillations within the ellipses which intersect
      the $x$-axis at $x = \pm m$ and
      $ x = \pm(\frac 83 F_\pi^2\mu^2/\Sigma -m)$.
      The labels N, $\pi$, K refer to the
      normal, pion and kaon condensed phases of the generating
      functional of the eigenvalue density.}
  \end{center}
\end{figure}

The extent of the oscillating region in the complex eigenvalue 
plane can be
understood from a mean field treatment of the chiral Lagrangian for the
generating functionals of the eigenvalue density (see \cite{osvphase}). 
The picture that emerges
is shown in Fig. \ref{fig:rhoPD} where we present results for one dynamical 
flavor. Inside the green ellipses the eigenvalue density is
strongly oscillating. Outside these regions the oscillating terms are 
exponentially suppressed and the eigenvalue density approaches the quenched
spectral density. The spectral density can be evaluated by  
introducing valence quark
masses $z$ and $z^*$ in addition to the dynamical quark with mass $m$. The
first two masses can be thought of as the up and down quark whereas the mass 
$m$ can be interpreted as a strange quark mass. Varying the chemical potential
then corresponds to varying the isospin chemical potential and the 
strangeness chemical potential explaining the appearance of regions we
identify as the pion condensed phase (denoted by $\pi$), 
the kaon condensed phase (denoted by $K$) and the normal phase 
(denoted by $N$). In the latter phase the chiral condensate is nonzero but
the Bose condensates are vanishing.  For more details we refer to
\cite{osvphase}. 

The fact that the structure of the eigenvalue density is determined by the
phases of the generating functional shows that this structure 
persists beyond the microscopic domain. In this paper, however,
we will focus on the microscopic limit. In this limit the QCD partition function
and the spectral density of the Dirac operator 
are uniquely determined by the global symmetries of QCD and 
hence  can be derived  from chiral random matrix theories with the same
symmetries \cite{V}. In random matrix theory, the microscopic limit 
can be taken by identifying the size of the matrices with  the volume of
space-time and defining  a scaling limit as in (\ref{micros}). A random matrix
model for the microscopic domain of QCD at nonzero chemical potential 
amenable to an analysis by orthogonal polynomials was introduced in \cite{O}, 
and its properties were further analyzed in \cite{AOSV} and \cite{OSV}.

The purpose of the present paper is two-fold. First we will show that 
within the random matrix model \cite{O} the delicate cancellations follow
from orthogonality properties of orthogonal polynomials. The
results are exact, independent of the size of the matrices. It allows
us to establish the $\mu$-independence of the full microscopic chiral 
condensate as obtained from integrating
the microscopic spectral density according to Eq. 
(\ref{rhoSigma}). We also  
present a direct computation using complex contour integrations which 
establishes the relation between the exact $\mu$-dependent
microscopic spectral density and the $\mu$-independent chiral condensate.  
As a by-product we will obtain the exact expressions for the partially
quenched chiral condensate at non-zero chemical potential as well as the
bosonic partition functions in an arbitrary fixed topological sector.  

In the upcoming section the random matrix model is introduced along
with the complex orthogonal polynomials and their relation, through
the Cauchy transform and kernels, with the eigenvalue density.  In
section \ref{sec:OSV} we then use the orthogonal polynomials to
establish the $\mu$-independence of the chiral condensate. A strategy
to derive this result without the use of orthogonal polynomials is
sketched in section \ref{sec:totalD}, but it could only be worked out
in detail in the strong nonhermiticity limit (see \ref{sec:appB}).  In
section \ref{sec:bosPF} we turn to the bosonic partition function and
the quenched and partially quenched condensates are discussed in
section \ref{sec:PQ} .  In the appendices we provide alternate
derivations of the results of section \ref{sec:OSV}.
\ref{sec:directmicro} reproduces the microscopic condensate starting
from the microscopic density so that only universal quantities are
used in the calculation.  \ref{sec:contour} uses a method involving
complex contour integration while in \ref{sec:appB} we first rewrite
the eigenvalue density as a total derivative.

\section{The random matrix model}
\label{sec:rmt}

We consider the microscopic limit of QCD at nonzero chemical
potential. In this limit where ($\Sigma$ is the magnitude of the
chiral condensate in the chiral limit and $F_\pi$ is the tree level
pion decay constant)
\be
 m\Sigma V \qquad {\rm and }\qquad \mu^2 F_\pi^2 V
\ee
are kept fixed for $V\to \infty$, the mass and chemical potential
dependence of the QCD partition function is given by a random matrix
model (see for example the review \cite{Tilo-Jac}). For $N_f$ quark
flavors with mass $m$ and $n$ pairs of regular and conjugate quarks
with masses $x$ and $y^*$, respectively, this partition function is
defined by \cite{O}
\be
 {\cal Z}_N^{N_f,n}(m,x,y^*;\mu) &\equiv& \int
 d\Phi d\Psi \ w_G(\Phi) w_G(\Psi)
 {\det}^{N_f}(\,{\cal D}(\mu) + m\,) \  \nn \\ 
 &&\times {\det}^n(\,{\cal D}(\mu) + x\,){\det}^n(\,{\cal
  D}^\dagger(\mu) + y^* ) ,
\label{ZNfNb}
\ee
where the non-Hermitian Dirac operator is given by
\be
\label{dnew}
\mathcal{D}(\mu) = \left( \begin{array}{cc}
0 & i \Phi/\alpha + \mu \Psi \gamma/\alpha \\
i \Phi^{\dagger}/\alpha + \mu \Psi^{\dagger} \gamma/\alpha & 0
\end{array} \right) ~.
\ee
Here $\Phi$ and $\Psi$ are complex $(N+\nu)\times N$ matrices both distributed
according to a Gaussian weight function 
\be
\label{wg}
w_G(X) ~=~ \exp( \, - \, N \, \tr \, X^{\dagger} X \, ) ~.
\ee
The parameters $\alpha$ and $\gamma$ are scale factors used to map the
random matrix model onto the chiral Lagrangian, which are given below.
The number of additional rows as compared to columns gives rise to $\nu$
zero modes of $\cal D$, and $\nu$ is therefore referred to as the topological
index.    
Inverse determinants are interpreted as bosonic quarks and will be denoted
by negative values of $N_f$ or $n$.  In cases where $N_f$ or $n$ are
zero they will be left off along with the corresponding masses.

In the random matrix model, the microscopic limit is given by the limit
$N\to \infty$ where
\be
\hat{m} = 2 N \alpha m  \ \ \ {\rm and} \ \ \ \hat{\mu}^2=2N \gamma^2 \mu^2 
\ee    
are kept fixed as $N\to\infty$. The identification with the QCD partition 
function is made according to 
(see the discussion in \cite{AOSV}) 
\be\label{mapping}
\hat{m}= 2 N \alpha m & \to &  m \Sigma V, \\
\hat{\mu}^2 = 2 N \gamma^2 \mu^2 & \to &  \mu^2 F_\pi^2 V. \nn
\ee
This then determines the scale factors $\alpha = \Sigma V/2N$ and
$\gamma^2 = F_\pi^2 V/2N$.  From here on we will drop the factors of $\alpha$
and $\gamma$ except when explicitly needed.

Contrary to the Hermitian
random matrix ensembles, it is quite nontrivial to express
the partition function (\ref{ZNfNb}) as an integral over
the joint probability distribution of the eigenvalues of 
$\cD(\mu)$. Remarkably, 
it was shown in \cite{O} that an analytical form could be obtained with
result given by
\be
\label{epfnew}
{\cal Z}_N^{N_f,n}(m,x,y^*;\mu)  \sim
m^{\nu N_f }(xy^*)^{n\nu} \int_{\mathbb{C}} \prod_{k=1}^{N} d^2z_k \,
{\cal P}^{N_f,n}(\{z_i\},\{z_i^*\}, m, x, \conj{y}; \mu),
\ee
where the  integration extends over the complex plane and 
the joint probability distribution reads
\be
\label{jpd}
{\cal P}^{N_f,n}(\{z_i\},\{z_i^*\}, m, x, \conj{y};\mu)
&=&  \frac{1}{\mu^{2N}} \left|\Delta_N(\{z_l^2\})\right|^2 \, 
\prod_{k=1}^{N} w(z_k,z_k^*;\mu) (m^2-z^2_k )^{N_f} (x^2-z_k^2 )^{n}
(y^{*\,2}-z_k^{*\,2} )^{n} .\nn\\
\ee 
The Vandermonde determinant is defined as  
\be
\Delta_N(\{z^2_l\}) \equiv \prod_{i>j=1}^N (z_i^2-z_j^2),
\label{vander}
\ee
and the weight function includes a modified Bessel function,
\be 
w(z_k,z^*_k;\mu) &=& |z_k|^{2\nu+2} 
K_\nu \left( \frac{N (1+\mu^2)}{2 \mu^2} |z_k|^2 \right)
\exp\left(-\frac{N (1-\mu^2)}{4 \mu^2}  
(z^2_k + \conj{z_k}^2) \right). 
\label{wnew}
\ee
The modified Bessel function is obtained as a result of the integration
over the angular degrees of freedom. In the microscopic limit the same 
functional dependence is obtained starting from a chiral Lagrangian 
for the bosonic phase quenched partition function \cite{SplitVerb2}. This 
is a strong argument for the universality of this factor.

The eigenvalue representation makes it possible to evaluate integrals 
over eigenvalues by means of the method of complex 
orthogonal polynomials \cite{A03,AV,BI,BII,AP}. Analytical
expressions for the eigenvalue density \cite{O}, 
eigenvalue correlation functions \cite{O} and partition functions 
\cite{AOSV,SVbos} have been obtained using this method.

\subsection{Orthogonal polynomials and their Cauchy transform}

In this section we introduce the orthogonal polynomials and relate them to
the quenched and unquenched eigenvalue density.
The complex orthogonal
polynomials corresponding to the weight function (\ref{wnew}) 
are given in terms of the complex Laguerre polynomials by \cite{O}
\be
p_k(z;\mu) = \left( \frac{1-\mu^2}N\right )^k k! 
L_k^\nu \left ( -\frac{Nz^2}{1-\mu^2} \right).
\ee 
They satisfy the orthogonality relations
\be
\int_{\mathbb{C}}d^2z\ w(z,z^*;\mu)\ p_k(z;\mu)\ p_l(z;\mu)^* ~ 
 ~=~ \delta_{kl} ~ r_k^\nu ~,
\label{Jo1}
\ee
with the norm $r_k^\nu $ given by 
\be
\label{Norm}
r_k^\nu ~=~
\frac{  \pi \, \mu^2 ~ (1+\mu^2)^{2k+\nu} ~ k! ~ (k+\nu)!}
     {N^{2k +  2+\nu}}  ~.
\ee
Since the orthogonal polynomials $p_k$ are related to the Laguerre polynomials
they also satisfy an orthogonality relation on the imaginary axis,
\be
\int_{0}^{\infty} dx ~ w_i(x) ~ p_k(i x;\mu) ~ p_l(i x;\mu)
 ~=~ \delta_{kl} ~ s_k^{\nu}  , 
\label{opreal}
\ee
with
\be
w_i(x) = x^{2\nu+1} \e^{-N x^2/(1-\mu^2)} \qquad {\rm and }\qquad
s_k^{\nu} = \frac{k! (k+\nu)! (1-\mu^2)^{2k+\nu+1}}{ 2 N^{2k+\nu+1}}  .
\ee
We will make use of this relation in section \ref{subsec:Sigmak}. Since
\be
\Pi_{l-1} = \frac{p_l(z)-p_l(m)}{z^2-m^2}
\ee
is a polynomial of order $l-1$ in $z^2$, as a direct consequence
of the orthogonality relations, we can establish the identity
\be
\int_\mathbb{C} d^2 z~ w(z,\zs;\mu) p_l(\zs)\frac{p_l(z)-p_l(m)}{z^2-m^2} = 0.
\label{J1}
\ee 

The Cauchy transform of the orthogonal polynomials is defined as
\be
\label{cauchy}
h_k(m;\mu) = \int_{\mathbb{C}} d^2z \frac 1{z^2-m^2}w(z,z^*;\mu) p^*_k(z;\mu),
\ee
where we recall that ${\mathbb{C}}$ indicates that the integration extends
over the complex plane.

The partition function for one fermion can be expressed in terms
orthogonal polynomials as
\be
\label{ZNf1}
\frac{Z_N^{N_f=1}(z;\mu)}{Z_N^{N_f=0}} = z^\nu p_N(z;\mu),
\ee
and the partition function for one bosonic flavor is given by a Cauchy 
transform \cite{AP,AOSV,SVbos}
\be
\frac{Z_N^{N_f=-1}(z;\mu)}{Z_N^{N_f=0}} = -\frac{z^{-\nu}
  h_{N-1}(z;\mu)}{r_{N-1}}. 
\label{ZbosFROMh}
\ee
Note that at finite $N$ both partition functions depend on the chemical
potential.
This dependence 
can be removed \cite{O-pq} from the fermionic partition function (\ref{ZNf1})
by 
scaling the factors $\alpha$ and $\gamma$ in
(\ref{dnew}), but we will not bother here since the $\mu$-dependence
will factorize in the microscopic limit which is our main
concern. The bosonic partition function has a nontrivial $\mu$-dependence that
can not be removed through the constants $\alpha$ and $\gamma$
\cite{SVbos,SVZ}. 
Below we will drop $\mu$ from the argument of $p$ and $h$ to make
the notation less clumsy.

\subsection{Kernels and the spectral densities}

~From the orthogonal polynomials and their Cauchy transforms we can construct
two different kernels (see for example \cite{AOSV}) for later use
\be
{\cal K}_N(x,y) &=& \sum_{k=0}^{N-1}\frac {p_k(x) p_k(y)}{r_k}, \\
{\cal H}_N(x,y) &=& \sum_{k=0}^{N}\frac {p_k(x) h_k(y)}{r_k}.
\ee
We will also use the auxiliary kernel
\be
{\cal N}_N(x,y) = \frac 1{y^2-x^2} + {\cal H}_N(x,y),
\label{Ndef}
\ee
which enters in the computation of the partially quenched chiral condensate
in section \ref{sec:PQ}.

\begin{figure}[t]
  \unitlength1.0cm
  \begin{center}
  \begin{picture}(3.0,2.0)
  \put(-5.,-5.){
  \psfig{file=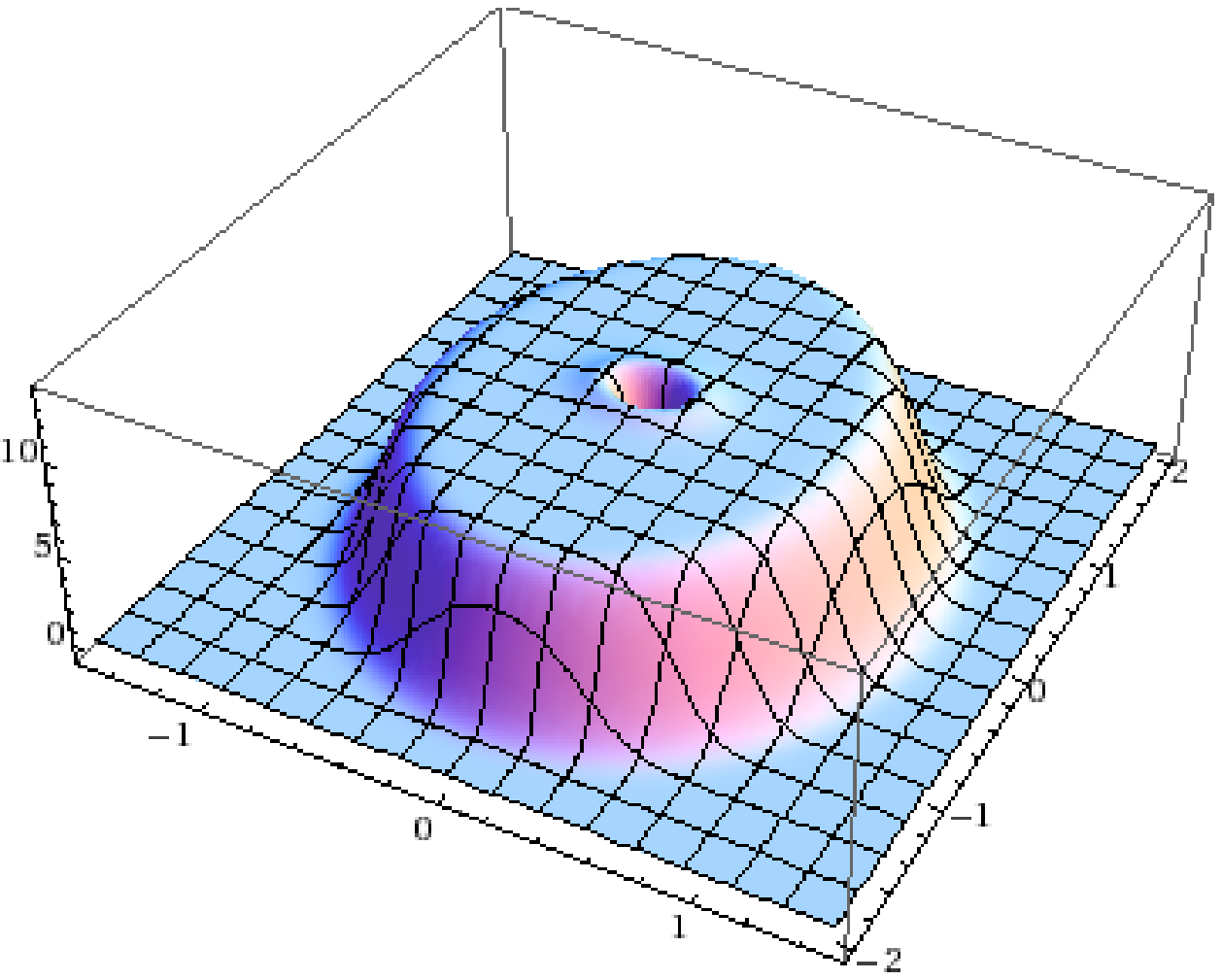,clip=,width=10cm}}
  \put(-1.5,-4.9){\bf\large Re$[z]$}
  \put(4.5,-2.0){\bf\large Im$[z]$}
  \put(-6.8,1.7){\bf \Large $\rho^{N_f=0}_{N}(z,z^*;\mu)$}
  \put(-5.,-14.){
  \psfig{file=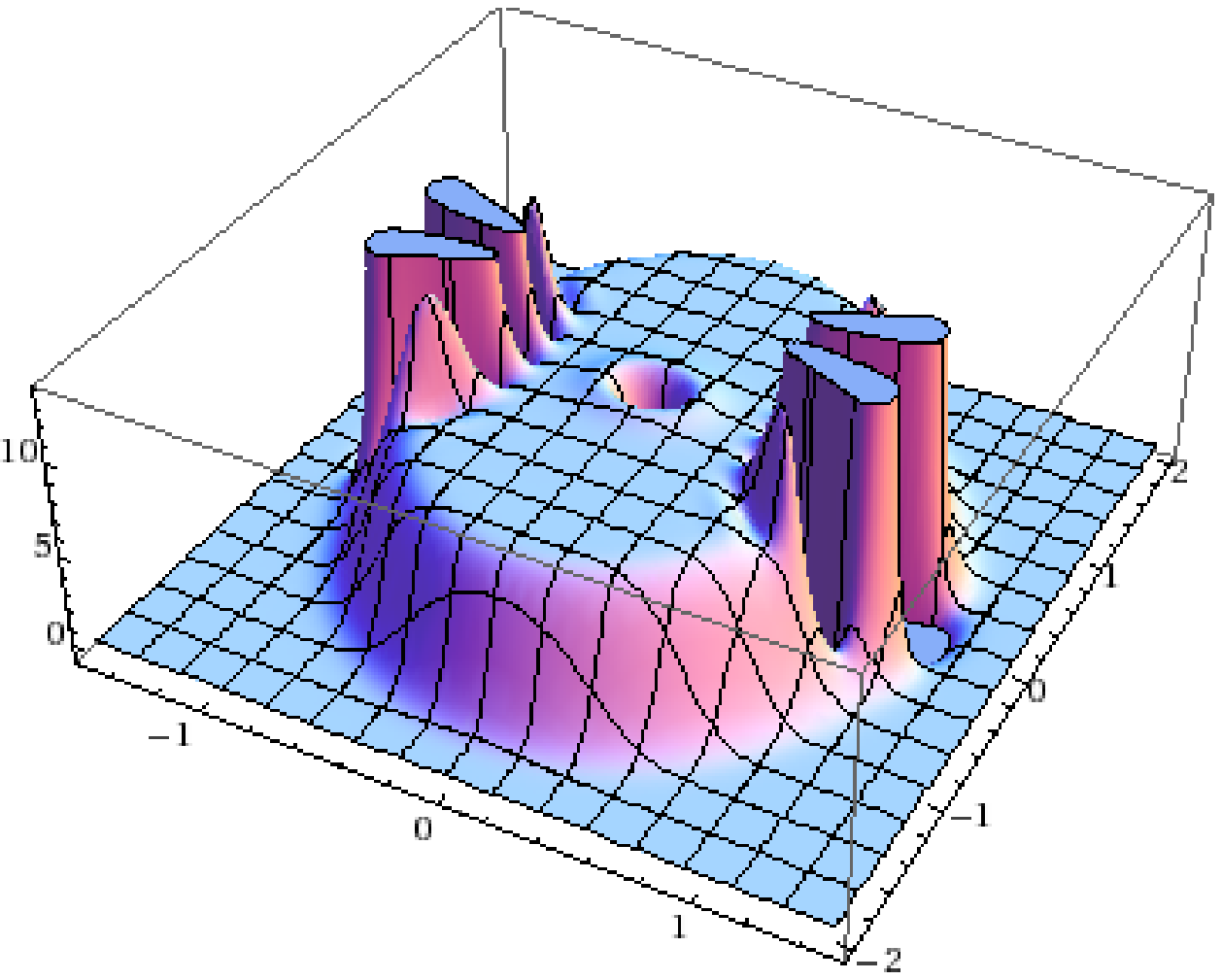,clip=,width=10cm}}
  \put(-1.3,-13.8){\bf\large Re$[z]$}
  \put(5,-10.4){\bf\large Im$[z]$}
  \put(-6.8,-7.){\bf \Large ${\rm Re}[\rho^{N_f=1}_{N}(z,z^*,m;\mu)]$}
  \end{picture}
  \vspace{14cm}
  \end{center}
\caption{\label{fig:rho} The eigenvalue density of the random matrix
  model (\ref{ZNfNb}) with $N=20$, $\mu=0.8$, $m=0.6$ and $\nu=0$.  
  The quenched density (top) is real and positive.
  The real part of the unquenched density (bottom) shows oscillations. 
Note that the oscillating region
  starts out at $z=\pm m$ and extends outward. The peak of the oscillations 
  has been cut; the maximum 
amplitude is an order of magnitude larger than the
  scale displayed.}
\end{figure}

The $\mu$-dependent spectral density is related to the kernel ${\cal K}$. In
particular, the quenched spectral density is given by \cite{O}
\be
\rho^Q_N(z,z^*;\mu) = 2 w(z,z^*; \mu) {\cal K}_N(z,z^*),
\ee
and the unquenched $N_f=1$ spectral density can be written as \cite{O}
\be
\rho^{N_f=1}_N(z,z^*,m;\mu) = 2 w(z,z^*;\mu)
 \left[ \cK_{N}(z,\zs) - \frac{p_N(z)}{p_N(m)} \cK_{N}(m,\zs) \right].
\label{rhoNf1}
\ee
Note that the first term is equal to the quenched spectral density.
The second term is responsible for the strong oscillations of the real 
and imaginary parts of the unquenched eigenvalue density (see figure
\ref{fig:rho}).   
There is also an extra contribution $\nu\delta^2(z)$ to the eigenvalue
density which  arises due to the exact zero modes of the Dirac operator which
has not been included in the spectral density.

\section{The $\mu$-independence of the microscopic condensate}
\label{sec:OSV}

In this section we will show how a $\mu$-independent chiral condensate
in the microscopic limit can arise even though the eigenvalue density of
Dirac operator which enters (\ref{rhoSigma}) has a strong $\mu$-dependence. 
Inserting the finite $N$ one flavor partition function (\ref{ZNf1})
in (\ref{SigmaFromZ}) we immediately find that the chiral condensate
is given by
\be
 V \Sigma(m) = \frac{d p_N(m)/dm}{p_N(m)}+\frac{\nu}{m}.
\ee
Making use of  orthogonality relations for the polynomials we will show 
that the above expression also follows from a direct integration of the
 spectral density (\ref{rhoNf1}). 
The microscopic result is then obtained by taking the microscopic limit 
of the finite N result, and as we shall see explicitly below, 
the $\mu$-dependence
drops out. The topological term $\nu/m$ follows trivially from the $\delta$
  function contribution due to the zero modes. 

Alternatively one can start from the microscopic limit of the spectral
density and show by explicit evaluation of the integrals that the
corresponding chiral condensate does not depend on the chemical
potential. For completeness, this derivation is given in
\ref{sec:directmicro}.  The techniques used here will also be useful
when calculating the partially quenched chiral condensate in section
\ref{sec:PQ}.

\subsection{Chiral condensate at finite $N$}

In this subsection we will calculate the one flavor 
chiral condensate by integrating
over the eigenvalue density (\ref{rhoNf1}) using a variant of (\ref{rhoSigma})
obtained from exploiting the evenness of $\rho$ in $z$
\be
V \Sigma(m) = \int_\mathbb{C} d^2z\frac{m}{m^2-z^2} \rho^{N_f=1}_N(z,\zs,m;\mu) .
\label{Sigma}
\ee
The spectral density for $N_f = 1$ given in Eq. (\ref{rhoNf1}) 
can be written out as
\be
\rho^{N_f=1}_N(z,z^*,m;\mu) =2 
w(z,z^*;\mu)\sum_{k=0}^{N-1}\frac{p_k(z^*)[p_k(z)-p_N(z)p_k(m)/p_N(m)]}{r_k} .
\label{spdens}
\ee

By subtraction and addition of  $p_k(m)$ and $p_N(m)$ to $p_k(z)$ and
$p_N(z)$, respectively, we obtain
\be 
\rho^{N_f=1}_N(z,z^*,m;\mu) =2
w(z,z^*;\mu)\left[\sum_{k=0}^{N-1}\frac{p_k(z^*)[p_{k}(z)-p_k(m)]}{r_k}
-\sum_{k=0}^{N-1} \frac{p_k(z^*)p_k(m)}{r_k} \frac{[p_N(z)-p_N(m)]}{p_N(m)} \right ].
\label{spec-dec}
\ee
The first sum in the square brackets is well behaved in the thermodynamic limit
whereas the second sum, with the exception of $z\approx m$,
shows oscillations with an amplitude that grows 
exponentially with the volume. 
The chiral condensate is given by the integral
\be
V \Sigma(m) = 2m \int_\mathbb{C} d^2 z ~ w(z,z^*;\mu) 
\left[\sum_{k=0}^{N-1}\frac{p_k(z^*)[p_{k}(z)-p_k(m)]}{r_k(m^2-z^2)}
-\sum_{k=0}^{N-1} \frac{p_k(z^*)p_k(m)}{r_k} \frac{[p_N(z)-p_N(m)]}{p_N(m)(m^2-z^2)} \right ].
\label{spec-dec1}
\ee
The contribution of the first sum in (\ref{spec-dec1})
integrates to zero due to orthogonality.  To obtain a result
that is  well behaved in the thermodynamic limit, 
the exponentially large terms in the second sum have
to be canceled which indeed happens because the
contributions to the chiral condensate are localized on $z=m$.
The reason is that, as a consequence of the orthogonality of the $p_k$, the sum
\be
w(z,z^*;\mu) \sum_{k=0}^{N-1} \frac{p_k(z^*)p_k(m)}{r_k} 
\label{sumk}
\ee
is a reproducing kernel (or equivalent to the delta function
$\delta^2(z-m)$) in the space of polynomials in $z$ of order
less than $N$ (denoted by $\Pi_n(z)$). We thus have
\be
\int_\mathbb{C} d^2 z ~ w(z,\zs;\mu)
\sum_{k=0}^{N-1} \frac{p_k(z^*)p_k(m)}{r_k} \Pi_{n}(z) = \Pi_{n}(m) 
~~~\mathrm{for} ~~ n < N.
\ee
Using the definition of derivative we finally find the following
result for the chiral condensate,
\be
V \Sigma(m) = \frac{2m}{p_N(m)}\lim_{z\to m} \frac {p_N(z)-p_N(m)}{z^2-m^2} 
= \frac{dp_N(m)/dm}{p_N(m)}. 
\label{SigmaN}
\ee
Clearly this is consistent with expressing the chiral condensate as 
the derivative with respect to $m$ of the logarithm of the partition function
(\ref{ZNf1}). Notice that since the spectral density (\ref{rhoNf1}) does not contain the
trivial contribution of the zero modes, $\nu\delta^2(z)$, 
the condensate calculated this way does not have 
the $\nu/m$ term. A second remark is that at finite $N$, the polynomial
$p_N(m)$ depends on the chemical potential. As will be shown in the next
subsection, this dependence drops out of the chiral condensate in the
microscopic limit.

\subsection{Microscopic limit of the finite $N$ results}

In this section we discuss the microscopic limit of the argument presented 
in the previous section. 
First, let us show explicitly that, in the microscopic limit, 
we obtain the expected $\mu$-independent chiral condensate. 
The microscopic limit of the orthogonal polynomials is given by
\be
 \frac{p_k(z;\mu)}{\sqrt{r_k}} \stackrel{N \to \infty}{\to}
 (\alpha z)^{-\nu}I_\nu(2N\alpha z t)
 \frac{Ne^{-2N \gamma^2 \mu^2 t^2 } }{\gamma \mu \sqrt \pi }
 \qquad {\rm with } \qquad t =\sqrt{\frac kN}.
\label{p-mikro}
\ee
Because of this factorization, 
the overall $\mu$-dependence of the orthogonal polynomials 
drops out of equation (\ref{SigmaN}) in the microscopic limit resulting in a
$\mu$-independent chiral condensate
\be
 \hat\Sigma(\hat{m}) \equiv
 \lim_{N\to\infty} \frac{\Sigma(\hat m/2N \alpha)}{2N \alpha} =
 \frac{d I_\nu(\hat{m})/d\hat m}{I_\nu(\hat{m})}-\frac{\nu}{\hat m}.
\ee  
The term $\nu/\hat m$ is canceled by the contribution of the zero modes.
In the above argument we have manipulated the non-universal finite $N$
expressions into a form which are suitable to take
the microscopic limit. In \ref{sec:directmicro} we will 
show that it is also possible to take the microscopic limit from the start and
deal with the universal expressions throughout the argument.

In fact, a slightly stronger result is valid. In the microscopic limit, 
orthogonal polynomials of the same order but with different arguments
have the same   $\mu$-dependence (see (\ref{p-mikro})).  Therefore, the ratio
\be
\frac{p_N(z)-p_N(m)}{p_N(m)(m^2-z^2)} = \sum_{k=0}^{N-1} c_k(m) 
\frac{p_k(z)}{p_N(m)}
\label{kernelexp}
\ee
does not depend on the chemical potential in the microscopic limit, and for
the same reason, each term in the sum does not depend on $\mu$.
If we define  $\Sigma_k$ as
\be
\Sigma_k(m) = \frac{2m p_k(m)}{p_N(m)} c_k(m),
\label{sigkm}
\ee
so that  
\be
\Sigma(m) = \sum_{k=0}^{N-1} \Sigma_k (m),
\ee
then also the $\Sigma_k$ are independent of $ \mu$ in the microscopic limit.
Explicit results for the microscopic limit of $\Sigma_k(m)$ will
be worked out in the next subsection.

\subsubsection{Microscopic limit of $\Sigma_k$}
\label{subsec:Sigmak}

In this subsection we derive the microscopic limit of $\Sigma_k(m)$ defined
in (\ref{sigkm}). This quantity, denoted by $\hat\Sigma(\hat m,t)$, also enters
in derivation of the chiral condensate by complex contour integration
(see \ref{sec:appB}).

The coefficient $c_k(m)$ defined in (\ref{kernelexp}) can be calculated by using
the orthogonality of the $p_k$ on the positive imaginary axis
\be\label{ck}
c_k(m) = \int_0^\infty dx w_i(x) \frac{p_k(ix)}{s_k^\nu}\frac{p_N(m)-p_N(ix)}{x^2+m^2}.
\ee
Now looking at the first part of the r.h.s. of the previous equation we define
\be
\Sigma_k^{\rm I}(m) = 2 m \frac{p_k(m)}{s_k^\nu}\int_0^\infty dx w_i(x)
\frac{p_k(ix)}{x^2+m^2} .
\ee
In the microscopic limit we have
\be
\sum_{k=0}^{N-1} \to \int_0^1 2Nt dt\qquad {\rm with } \qquad t = \sqrt{k/N}
\qquad  {\rm and} \qquad w_i(x) = x^{2\nu+1} ,
\ee
so that
\be
\hat\Sigma^{\rm I}(\hat m,t) = 2\hat m \, t \, \hat m^{-\nu}I_\nu(\hat m t) 
\int_0^\infty d\hat{x}\frac {\hat x^{\nu+1}} 
{\hat x^2 + \hat m^2} J_\nu (\hat x t) = 2\hat m t I_\nu(\hat m t) K_\nu(\hat m t).
\ee 
Likewise the second part of $c_k(m)$  contributes to the chiral condensate as
\be
\Sigma_k^{\rm II}(m) = 2 m \frac{p_k(m)}{p_N(m) s_k^\nu} \int_0^\infty dx w_i(x) \frac{p_k(ix) p_N(ix)}{x^2+m^2}.
\ee
In the microscopic limit this simplifies to
\be
\hat\Sigma^{\rm II}(\hat m,t) = 2\hat m t \frac {I_\nu(\hat m t)}{I_\nu(\hat m)}
\int_0^\infty d\hat x \frac {\hat x}{\hat{x}^2+\hat{m}^2} 
J_\nu(\hat x t) J_\nu(\hat x).
\ee
The integral is known \cite{Gradshtein} resulting in
\be 
\hat\Sigma^{\rm II}(\hat m,t) = 2\hat m t \frac {I_\nu(\hat m t)}{I_\nu(\hat m)}
K_\nu(\hat m) I_\nu(\hat m t).
\ee
We conclude that the microscopic limit of $\Sigma_k(m)$ is independent
of the chemical potential.

Reminding ourselves of the identities
\be
\int_0^1 dt t [I_\nu(\hat mt)]^2 &=& \frac 12[I^2_\nu(\hat m) 
- I_{\nu+1}(\hat m)I_{\nu-1}(\hat m)],\nn \\
\int_0^1 dt t K_\nu(\hat mt)I_\nu(\hat mt) &=& \frac 12 [K_\nu(\hat m)I_\nu(\hat m)+K_{\nu+1}(\hat m)I_{\nu -1}(\hat m)] -\frac \nu{\hat m^2},
\ee
we obtain after using the Wronskian identity
\be
\hat\Sigma(\hat m) = \int_0^1 dt [\hat\Sigma^{\rm I}(\hat m, t) 
-\hat\Sigma^{\rm II}(\hat m, t) ] =\frac{I_\nu'(\hat m)}{I_\nu(\hat m)}
-\frac{\nu}{\hat m},
\ee
which is the correct result after adding the contribution of the zero modes.

We have by now established the $\mu$-independence of the microscopic chiral
condensate as expressed by an integral over the eigenvalue density. As
discussed next, the properties of the
spectral density that lead to the $\mu$-independence of the 
chiral condensate can be exposed further by writing the 
the spectral density as a derivative of the partially quenched chiral
condensate.

\section{The spectral density as a total derivative}
\label{sec:totalD}

In section \ref{sec:OSV} we have seen that the chiral condensate can
be written as an integral over a reproducing kernel in the space
of complex orthogonal polynomials up to order $N-1$:
\be
\Sigma(m) = -2m\int_\mathbb{C} d^2 z K_{N}(z^*, m) \frac {\Pi_{N-1}(z)}{p_N(m)} .
\ee
In the thermodynamic limit the kernel can be written as
\be
\lim_{N\to\infty}K_{N}(z^*,m)  \sim 2m \delta^2(z^2-m^2) 
= \frac 1 \pi \del_{z^* } \frac 1{z^2-m^2}.
\ee
Using a partial integration this suggest that the condensate can be written as
\be
\Sigma(m) = \int_\mathbb{C} d^2 z \frac {2m}{z^2 - m^2} \del_{z^*} F(z,z^*,m;\mu),
\label{F}
\ee
or that the spectral density can be written as a total derivative with respect
to $z^*$. Such a representation of the eigenvalue density in the complex
plane is well known \cite{Girko,Misha}; the function $F$ 
is known as the partially quenched condensate
\be
 F(z,z^*,m;\mu) = 
 \frac 1V
\left\langle \Tr \frac{1}{\dslash+\mu\gamma_0+z}\right\rangle.
\ee
It is the chiral condensate evaluated at a complex mass $z$. A partial
integration of (\ref{F}) singles out $z=m$ and consistently reproduces the
chiral condensate $\Sigma(m)=F(z=m,z^*=m,m;\mu)$. 

If we can express the eigenvalue density as a total derivative 
$\rho = \del_{z^*} F(z,z^*,m;\mu)$ and show that $F$ at $z=z^*=m$ is
independent of $\mu$ we have an alternative way to show how a strongly
$\mu$-dependent density can result in a chiral 
condensate that does not depend on the chemical potential. 
This alternative proof will be worked out in \ref{sec:appB} 
in the strong
non-Hermiticity limit for arbitrary $\nu$.  
While the exact expressions for the microscopic eigenvalue density are known,
the partially quenched condensates at nonzero chemical potential have not
been obtained previously and will be derived in section \ref{sec:PQ}.

\section{Cauchy transform and the bosonic partition function}
\label{sec:bosPF}

In this section we evaluate the Cauchy transform of the orthogonal polynomials
for finite $N$. As a new result we obtain 
the microscopic limit of
the bosonic partition function with arbitrary topological charge. 
For zero topological charge the result agrees with the bosonic partition
function obtained in \cite{SVbos} and thus proves 
a conjectured cancellation in the derivation of \cite{SVbos}. The result of
this section can also be used to obtain an alternate expression for
the microscopic limit of $\cH_N(m_1,m)$ at finite $N$.  

The Cauchy transform is defined by (see Eq. (\ref{cauchy}))
\be
\label{hk}
h_k(m) &=&  \int_\mathbb{C} d^2z \; w(z, z^*; \mu) \frac{1}{z^2-m^2} p_k(\conj{z}) \el
&=& \int_\mathbb{C} d^2z \; \frac{|z|^{2\nu+2}}{z^2-m^2} K_\nu(b|z|^2)
\e^{-a(z^2+\zssq)} p_k(\zs)
\ee
with $a = N (1-\mu^2)/4\mu^2$ and $b = N (1+\mu^2)/2\mu^2$.
The integrals can be evaluated by writing
\be
\frac{\e^{-a z^2}}{z^2-m^2} &=& \frac{\e^{-a m^2}}{z^2-m^2} +
 \frac{\e^{-a z^2}-\e^{-a m^2}}{z^2-m^2} ~.
\ee
The first term has a pole while the second is analytic.
The angular integration for the first term is now easy since the integrand
only depends on $\zs$ (or $|z|$) and not $z$. Simply expanding $1/(z^2-m^2)$ 
in a geometric series in $z^2/m^2$ we obtain
\be\label{pole-term}
\int_\mathbb{C} d^2z \; \frac{|z|^{2\nu+2}}{z^2-m^2} K_\nu(b|z|^2)
 \e^{-a(m^2+\zssq)} p_k(\zs) 
= -\frac{2 \pi}{m^2} \int_0^m dr r^{2\nu+3} K_\nu(b r^2)
 \e^{-a(m^2+r^4/m^2)} p_k(r^2/m) \nonumber ~.
\ee
For the second piece, since it is analytic, we assume it can be expanded in
the form
\be
\label{exp}
\frac{\e^{-a z^2}-\e^{-a m^2}}{z^2-m^2} &=&
 \e^{-a z^2} \sum_{k=0}^{\infty} d_k(m) p_k(z) ~.
\ee
Then by substituting this into (\ref{hk}) and using orthogonality we get
\be
\int_\mathbb{C} d^2z \; |z|^{2\nu+2}
 \frac{\e^{-a z^2}-\e^{-a m^2}}{z^2-m^2} K_\nu(b|z|^2)
 \e^{-a \zssq} p_k(\zs) &=& r_k^\nu d_k(m),
\ee
where $r_k^\nu$ is the normalization factor defined in (\ref{Norm}).
To evaluate $d_k(m)$ we first assume that the series (\ref{exp}) 
converges for all $z$ and $m$.
Then this is also valid along the imaginary axis $z = i x$
where we can use the orthogonality relations of the orthogonal polynomials
$p_k(i x)$ defined in (\ref{opreal}).
The coefficients are thus given by
\be
d_k(m) = \frac{1}{s_k^{\nu}} \int_{x=0}^{\infty} dx w_i(x) p_k(i x)
\frac{\e^{-a(x^2+m^2)}-1}{x^2+m^2}.
\ee
Using that
\be
\int_0^a dt ~ \e^{-t(x^2+m^2)} = - \frac{\e^{-a(x^2+m^2)}-1}{x^2+m^2} ~,
\ee
and the formula \cite{Gradshtein}
\be
\int_0^\infty e^{-st}t^\nu L_k^\nu(t) dt = \frac {\Gamma(\nu+k+1)(s-1)^k}
{k!s^{k+\nu+1}}
\ee
this can  be reduced to 
\be
\label{sk}
d_k(m) =
 -\frac{ N^{k+1}}{k!(1-\mu^2)^{k+1}}
\int_{0}^{(1-\mu^2)^2/4\mu^2} ~
 \e^{-m^2N t/(1-\mu^2)} \frac{t^k}{(t+1)^{k+\nu+1}} ~ dt~.
\ee
Combining the above expressions we thus find 
\be
h_k(m) &=&-\frac{2 \pi}{m^2} \int_0^m dr r^{2\nu+3} K_\nu(b r^2)
 \e^{-a(m^2+r^4/m^2)} p_k(r^2/m) 
\nn\\ &&-
\frac{r_k^\nu N^{k+1}}{k!(1-\mu^2)^{k+1}} \int_{0}^{(1-\mu^2)^2/4\mu^2} dt ~
 \e^{-m^2N t/(1-\mu^2)} \frac{t^k}{(t+1)^{k+\nu+1}} ~.
\label{hkfin}
\ee
This form makes it easier to take the microscopic limit.

\subsection{Microscopic limit of the bosonic partition function }

The bosonic partition function given by (\ref{ZbosFROMh}) can now be obtained
using the expression (\ref{hkfin}) for the Cauchy transform. In
the microscopic limit we find (as in \cite{SVbos,signprd} a factor
$\exp(\hat\mu^2/2)$ has been removed to ensure independence of $\mu$ for
$\mu<m_\pi/2$ in the limit $\hat\mu\to\infty$) 
\be
Z^{(\nu)}_{N_f=-1}(\hat m;\hat \mu) &=& 
\frac{e^{-2\hat\mu^2-\hat m^2/8\hat \mu^2} 
e^N \sqrt{2 N}}{4 \hat \mu^2 \sqrt \pi}
\int_{0}^{\hat{m}} du u \exp[ -\frac {u^2}{8\hat\mu^2}]
K_\nu\left ( \frac{u\hat{m}}{4\hat\mu^2}\right )I_\nu(u)\nn \\
 && +
\frac{ e^ N \sqrt{2 N}}{\sqrt{\pi} (2\hat m)^\nu}
\frac 12 \int_0^{1/8\hat\mu^2}\frac {ds}{s^{\nu+1}}e^{-\hat m^2 s -1/4s}.
\label{OSVbos}
\ee

The microscopic limit of the $N_f=-1$ partition function 
in the sector of
zero topological charge was evaluated in \cite{SVbos,signprd}
with the result,
\be\label{Z-1_v1}
Z^{(\nu=0)}_{N_f=-1}&=&
e^{-2\hat{\mu}^2}\frac{1}{4\hat{\mu}^2} e^{-\frac{\hat{m}^2}{8\hat\mu^2}}
 \\ && \times \left [
\int_{0}^{\hat{m}} du u \exp[ -\frac {u^2}{8\hat\mu^2}]
K_0\left ( \frac{u\hat{m}}{4\hat\mu^2}\right )I_0(u) 
+\int_{0}^{\infty} du u \exp[ -\frac {u^2}{8\hat\mu^2}]
I_0\left 
( \frac{u\hat{m}}{4\hat\mu^2}\right ) K_0(u) \right ] . \nn  \\
&=& K_0(\hat{m})
-e^{-2\hat{\mu}^2}\frac{1}{4\hat{\mu}^2} 
e^{-\frac{\hat{m}^2}{8\hat\mu^2}} 
\int_{\hat{m}}^\infty du u\exp[ -\frac {u^2}{8\hat\mu^2}]
K_0\left ( \frac{u\hat{m}}{4\hat\mu^2}\right ) I_0(u). 
\label{Z-1_v2}
\ee
It is immediately clear that the first term agrees up to an overall
normalization constant with the first term in the result (\ref{OSVbos}).
That also the second terms are in agreement can be seen from the substitution
of 
\be
K_0(u) = \frac12 \int_0^\infty \frac{dt}{t} ~ \e^{-t-u^2/4t}
\ee
in (\ref{Z-1_v1}) which allows the $u$ integration to be performed
\be
\frac{1}{4\hat{\mu}^2} e^{-2\hat{\mu}^2-\frac{\hat{m}^2}{8\hat\mu^2}}\int_{0}^{\infty} du u \exp[ -\frac {u^2}{8\hat\mu^2}]
I_0\left( \frac{u\hat{m}}{4\hat\mu^2}\right ) K_0(u) 
&=& \frac{1}{2}\int_0^{1/8\hat\mu^2}\frac{dt}{t}e^{-\hat{m}^2t-1/4t}.\nn
\ee

\section{The partially quenched chiral condensate}
\label{sec:PQ}

The partially quenched chiral condensate for $N_f$ flavors is defined by
\be
\Sigma_{N_f}(m) &=& \frac1V \left[ \partial_J \ln
\left(\cZ_N^{N_f,N_b=1}(m_1, \cdots, m_{N_f}, m+J|m)\right)
 \right]_{J=0} .
\ee
In this section we evaluate  the 
microscopic limit of this expression for $N_f=0$ and $N_f=1$ and 
derive simplified expressions valid in the strong non-Hermiticity limit.    

For the quenched theory ($N_f=0$) we need the ratio \cite{BII}
\be
\frac{\cZ_N^{N_f=1,N_b=1}(x|y;\mu)}{\cZ_N^{N_f=0}} &\equiv&
 \left\langle\ \left(\frac{x}{y}\right)^\nu\prod_{j=1}^N
 \frac{(x^2-z_j^2 )}{(y^2-z_j^{2})}~\right\rangle_{N_f=0} \nn \\
 &=& ~ (y^2-x^2)\ \left(\frac{x}{y}\right)^\nu \cN_{N-1}(x,y),
 \label{NZ}
\ee
so that the quenched chiral condensate can be expressed as
\be
\label{sigmaQH}
V \Sigma_Q(m) = \frac{\nu}{m} - 2 m \cH_{N-1}(m,m) ~,
\ee
where we used the auxiliary kernel defined in (\ref{Ndef}). Below we will
evaluate the microscopic limit of $\cH_{N-1}(m,m)$.

What enters in the chiral condensate for $N_f =1$ is the ratio  
\be\label{Z2|1}
\frac{\cZ_N^{N_f=2,N_b=1}(x,m_1|y;\mu)}{\cZ_N^{N_f=1}(m_1;\mu)} &\equiv&
 \frac 1{\cZ_N^{N_f=1}(m_1;\mu)}
\left\langle\ \left(\frac{x m_1}{y}\right)^\nu\prod_{j=1}^N
 \frac{(x^2-z_j^2)(m_1^2-z_j^2)}{y^2-z_j^2}~\right\rangle_{N_f=0}  \\
 &=& ~ \frac 1{m_1^\nu p_{N}(m_1)}\left(\frac{x m_1}{y}\right)^\nu
 \frac{(y^2-x^2)(y^2-m_1^2)}{x^2-m_1^2}
 \left| \begin{array}{cc}
  \cN_{N}(x,y)   & p_N(x) \\
  \cN_{N}(m_1,y) & p_N(m_1) \\
 \end{array} \right| ~ ,\nn
\ee
where the second equality follows from \cite{BII}. The partially quenched
condensate for $N_f=1$ is the condensate as a function 
of $m$ for fixed physical mass $m_1$. It can be expressed as
\be
\label{sigmaPQ}
V \Sigma_{PQ}(m,m_1) = \frac{\nu}{m} - 2 m \cH_N(m,m) +\frac{2m}{m_1^2-m^2}
+2m\frac{p_N(m)}{p_N(m_1)}\left[\frac{1}{m^2-m_1^2}+\cH_N(m_1,m)\right] ~.
\ee
This form makes it easy to take the $m_1=m$ limit.
The extension to more flavors again only requires the function
$\cH_N(m_1,m)$. 

The integral appearing in the bosonic partition function (\ref{Z-1_v2}) 
has an essential singularity at $\mu = 0$ \cite{signprd}. The function $\cH$
and hence the  
quenched (\ref{sigmaQH}) and partially quenched (\ref{sigmaPQ}) condensate 
inherit this non-analyticity. 
For $m_1\neq m$ the 
partially quenched condensate (see (\ref{sigmaPQ})) is a function 
of the chemical potential.
However, for  $m_1= m$ it correctly reduces to the derivative with respect to
 $\hat{m}$   
of $Z_{N_f=1}$ which is $\hat\mu$-independent in the microscopic limit.

\subsection{Microscopic limit of the $\cH$-kernel}

Using the definition of the Cauchy transform, the $\cH$-kernel can be
written as
\be
\label{cH}
\cH_{N}(m_1,m) &=& \sum_{k=0}^{N} p_k(m_1) h_k(m) / r_k \el
&=& \int_\mathbb{C} d^2z \; w(z,z^*;\mu) \frac{1}{z^2-m^2}
 \sum_{k=0}^{N} p_k(m_1) p_k(\conj{z}) / r_k ~.
\ee
To evaluate $\cH_N(m_1,m)$ we substitute 
\be
p_k(m_1) = (z^2-m^2)[c_{k-1}p_{k-1}(z)+c_{k-2}p_{k-2}(z)+\ldots] 
+ p_k(z m_1/m) 
\ee
into (\ref{cH}) to get
\be
\cH_{N}(m_1,m) &=& \int_\mathbb{C} d^2z \; w(z,z^*;\mu) \frac{1}{z^2-m^2}
 \sum_{k=0}^{N} p_k(z m_1/m) p_k(\conj{z}) / r_k ~.
\ee
In the microscopic limit the $\cH$ kernel gets multiplied by
an additional factor $1/(2N)^2$. Using expressions for the microscopic
limit of the orthogonal polynomials we then obtain
\be
\label{cHmicro}
\hat{\cH}(\hat{m_1},\hat{m}) = \int_\mathbb{C} d^2\hat{z} \;
 \frac{|\hz|^2\hm^\nu}{4\pi\hmu^2\hm_1^\nu}
 K_\nu\left(\frac{|\hz|^2}{4\hmu^2}\right)
 \exp\left(-\frac{\hz^2+\conjsq{\hz}}{8\hmu^2}\right)
 \frac{1}{\hat{z}^2-\hat{m}^2} \el
 \times \int_0^1 tdt \; \exp(-2\hat{\mu}^2 t^2)
 I_\nu(\hat z {t} \hm_1/\hm) I_\nu(\conj{\hat{z}}{t}) ~.
\ee

\subsection{Strong non-Hermiticity limit}

The expression for $\cH$ simplifies considerably in the strong non-Hermiticity
limit, $\hat\mu\gg1$. Then for $\hat m_1 < \hat m$ using (\ref{largemu-app})
we obtain
\be
\hat{\cH}(\hat{m_1},\hat{m}) = \int_\mathbb{C} d^2\hat{z} \;
 \frac{|\hz|^2\hm^\nu}{16\pi\hmu^4\hm_1^\nu}
 K_\nu\left(\frac{|\hz|^2}{4\hmu^2}\right)
 I_\nu\left(\frac{|\hz|^2\hm_1}{4\hmu^2\hm}\right)
 \exp\left(\frac{\hm_1^2-\hm^2}{8\hmu^2\hm^2} \hz^2\right)
 \frac{1}{\hat{z}^2-\hat{m}^2} ~.
\ee
This is now in a form that can be evaluated analytically. The angular integration gives
\be
\hat{\cH}(\hat{m_1},\hat{m}) &=&
 \frac{\hm^{\nu-2}}{8\hmu^4\hm_1^\nu}
 \exp\left(\frac{\hm_1^2-\hm^2}{8\hmu^2}\right)
 \int_{\hm}^{\infty} 
  K_\nu\left(\frac{r^2}{4\hmu^2}\right)
  I_\nu\left(\frac{r^2\hm_1}{4\hmu^2\hm}\right)
  r^3 dr \el
&-&\frac{\hm^{\nu-2}}{8\hmu^4\hm_1^\nu}
 \int_{0}^{\infty} 
  K_\nu\left(\frac{r^2}{4\hmu^2}\right)
  I_\nu\left(\frac{r^2\hm_1}{4\hmu^2\hm}\right)
  r^3 dr ~.
\ee
The $r$-integrals are known \cite{Gradshtein} and the final result is
\be
\hat{\cH}(\hat{m_1},\hat{m}) &=&
 \frac{-1}{\hm^2-\hm_1^2} +
 \frac{\hm^{\nu+1}}{4\hmu^2\hm_1^\nu(\hm^2-\hm_1^2)}
 \exp\left(\frac{\hm_1^2-\hm^2}{8\hmu^2}\right) \times \el
 &&\left[
  \hm K_{\nu-1}\left(\frac{\hm^2}{4\hmu^2}\right)
   I_{\nu}\left(\frac{\hm\hm_1}{4\hmu^2}\right) +
  \hm_1 K_{\nu}\left(\frac{\hm^2}{4\hmu^2}\right)
   I_{\nu-1}\left(\frac{\hm\hm_1}{4\hmu^2}\right)
 \right] ~.
\label{hstrong}
\ee
Inserting this result in (\ref{sigmaQH}) and (\ref{sigmaPQ}) gives  
respectively the quenched and the partially quenched chiral condensate in the
strong non-Hermiticity limit. In both cases we need $\hat{\cH}$
at equal masses. With help of the Wronskian identity this can be expressed as
\be\label{hstrong-eqmass}
2\hat{m}\hat{\cH}(\hm,\hm) &=&
\frac{\nu}{\hat m}-\frac{\hm}{4\hmu^2}-\frac{\hm}{4\hmu^2}\left[
\nu\left(I_{\nu-1}(\frac{\hm^2}{4\hmu^2})K_{\nu}(\frac{\hm^2}{4\hmu^2})-I_{\nu}(\frac{\hm^2}{4\hmu^2})K_{\nu-1}(\frac{\hm^2}{4\hmu^2})\right)\right.
\\
& & \hspace{3cm}
\left.+\frac{\hm^2}{4\hmu^2}\left(I_{\nu-1}(\frac{\hm^2}{4\hmu^2})K_{\nu-1}(\frac{\hm^2}{4\hmu^2})+I_{\nu}(\frac{\hm^2}{4\hmu^2})K_{\nu}(\frac{\hm^2}{4\hmu^2})\right)\right].
\nn
\ee  

In the strong non-Hermiticity limit, the partially quenched chiral
condensate also follows from the expression of the spectral density as
a total derivative. In the quenched case we find
\be
\hat\Sigma_Q(\hat m) = \int_\mathbb{C} d^2 \hat z \frac 1{\hat z+\hat m}  
\frac 1\pi\del_{\hat z^*} F(\hat z, \hat z^*, \hat m) 
+ \frac \nu {\hat m}= -F(\hat z =\hat m, \hat z^*=\hat m,\hat m)
+ \frac \nu {\hat m},
\label{condq}
\ee
with $F$ given by (see (\ref{rhoqsub}) in \ref{sec:appB})
\be 
F(\hat z, \hat z^*, \hat m) =  \left . 
\frac {\del_{\hat m}[ F_A(\hat z, \hat m) G(\hat m \hat z, \hat z \hat z^*)]}
{2\hat z}\right|_{\hat m= \hat z}-
{\rm Sing}\left [\frac {\left . G(\hat z\hat m, \hat z \hat z^*)
\del_{\hat m} F_A(\hat z, \hat m)
\right|_{\hat m= \hat z}}{2 \hat z}\right ] .
\ee
The subtraction of the singular term amounts to not  differentiating
$I_\nu(\hat m)$ except for at $\hat m =0$ which is compensated for by subtracting
the term $\nu/{\hat m} $. To avoid convergence problems we
subtract the asymptotic value of the quenched spectral density which
is equal to $u$. The asymptotic contribution will be denoted by
$\hat\Sigma_{\rm as}(\hat m)$.
A simple calculation then results in
\be
\hat\Sigma_Q(\hat m) &=& \hat\Sigma_{\rm as}(\hat m) +
\frac{\hm}{4\hmu^2} \left[\nu
  K_{\nu+1}\left(\frac{\hm^2}{4\hat{\mu}^2}\right)I_\nu\left(\frac{\hm^2}{4\hat{\mu}^2}\right)- \nu K_\nu\left(\frac{\hm^2}{4\hat{\mu}^2}\right) I_{\nu+1}\left(\frac{\hm^2}{4\hat{\mu}^2}\right)\right. \nn\\
&& \left. + \frac{\hm^2}{4\hmu^2} \left [
K_\nu\left(\frac{\hm^2}{4\hat{\mu}^2}\right) I_\nu\left(\frac{\hm^2}{4\hat{\mu}^2}\right)
+K_{\nu+1}\left(\frac{\hm^2}{4\hat{\mu}^2}\right) I_{\nu+1}\left(\frac{\hm^2}{4\hat{\mu}^2}\right)\right ]-1\right]\nn \\
 &=& -\frac{\nu}{\hm} + \frac{\hm}{4\hmu^2} +
\frac{\hm^3}{16\hmu^4} \left[
  K_\nu\left(\frac{\hm^2}{4\hat{\mu}^2}\right)
  I_\nu\left(\frac{\hm^2}{4\hat{\mu}^2}\right) +
  K_{\nu+1}\left(\frac{\hm^2}{4\hat{\mu}^2}\right)
  I_{\nu-1}\left(\frac{\hm^2}{4\hat{\mu}^2}\right) \right] ~.
\label{SigmaQstrong}
\ee
This result can also be obtained by direct integration the
quenched spectral density using polar coordinates.
The contribution $\hat\Sigma_{\rm as}(\hat m)$  
is obtained by a direct calculation 
\be
\hat\Sigma_{\rm as}(\hat m) = \frac{1}{4\hmu^2} \int dx dy \frac 1{x+iy +\hat m} = 2\frac{\hm}{4\hmu^2} ,
\ee
where the $x$-integration has been taken over a symmetric interval about
$x=0$. As it should, the result (\ref{SigmaQstrong}) agrees 
with (\ref{sigmaQH}) after inserting (\ref{hstrong-eqmass}).  

\begin{figure}[t]
  \unitlength1.0cm
\centerline{  \epsfig{file=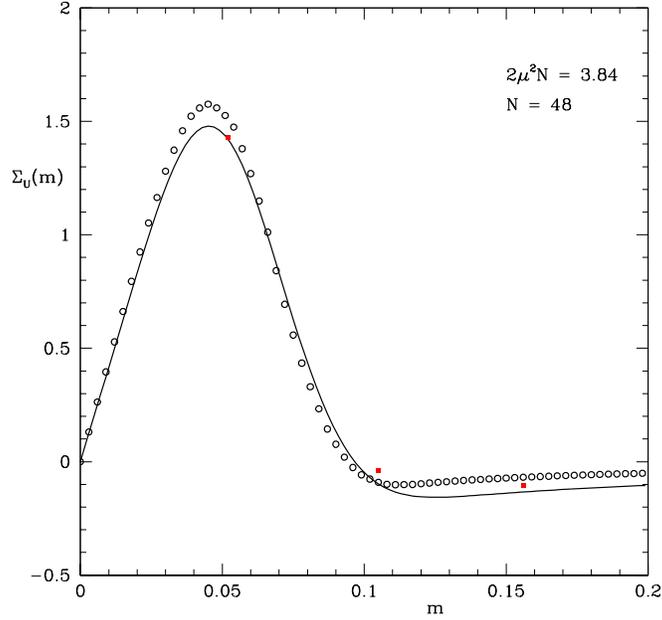,width=9cm}}
  \caption{The effect of unquenching on the partially
    quenched chiral condensate is shown by plotting  
$\Sigma_U(m)\equiv\hat\Sigma_{U,PQ}(\hat m_1=0,2N m)$, 
for one massless
flavor with  $\hat m_1=0$, $\nu = 0$ and $\hmu^2 =3.84$. 
The strong non-Hermiticity result is given by the
solid curve. The squares denote the exact microscopic result, and the
open circles show the result obtained in \cite{Halasz} 
by matrix diagonalization.}
\label{valmass}
\end{figure} 

Next we study the valence quark mass dependence of the partially quenched 
chiral condensate for $N_f = 1$ and compare with numerical simulations obtained in
\cite{Halasz} for $\nu=0$ and $m_1=0$. 
In this case the result for $\Sigma_{PQ}(m_1=0,m)$ is particularly 
simple. The chiral condensate is decomposed according to (\ref{qudec})
\be
\hat\Sigma_{PQ}(\hm_1=0,\hm) = \hat\Sigma_Q(\hm) - \hat\Sigma_{U,PQ}(\hm_1=0,\hm). 
\label{valanfor}
\ee
The quenched contribution was evaluated above (see (\ref{SigmaQstrong}))
and is given by (introducing the notation $u\equiv 1/4\hmu^2$)
\be
\hat\Sigma_Q(\hat m)&=& u\hat m+ u^2 \hat m^3[K_0(u\hat m^2)I_0(u\hat m^2) 
+ K_1(u\hat m^2)I_1(u\hat m^2)] .
\ee
Notice that this result is only valid when $\hat m$ is inside the domain of
eigenvalues.
The unquenched contribution follows immediately from (\ref{gensigu}) and
(\ref{genF}), or alternatively, can be calculated from the strong 
non-Hermiticity limit of $\cH(m,m_1) $ and (\ref{sigmaPQ}). The result
for $\nu =0$ and $\hat m_1 =0$ is given by
\be
\hat\Sigma_{U,PQ}(\hat{m}_1=0,\hat m)&=& -2u\hat m e^{-u\hat m^2/2} I_0(\hat m)
 K_1(u \hat m^2) + \frac 2{\hat m} .
\ee
The derivation of this result does not require that $\hat m$ is inside
the domain of eigenvalues as in the quenched case.

In Fig. \ref{valmass} we show the result for $\hat\Sigma_{U,PQ}(\hat{m}_1=0,\hat m)$ 
(solid curve) for $\hmu^2 = 3.84$. In this figure we also
give numerical results for an ensemble of $96\times 96$ matrices obtained
in \cite{Halasz} by direct diagonalization of the random matrices. 
 The solid red squares are the results obtained by numerical integration
of the exact result for the spectral density. 

\begin{figure}[t]
  \unitlength1.0cm
  \begin{center}
  \begin{picture}(3.0,2.0)
  \put(-5.,-6.){
  \psfig{file= 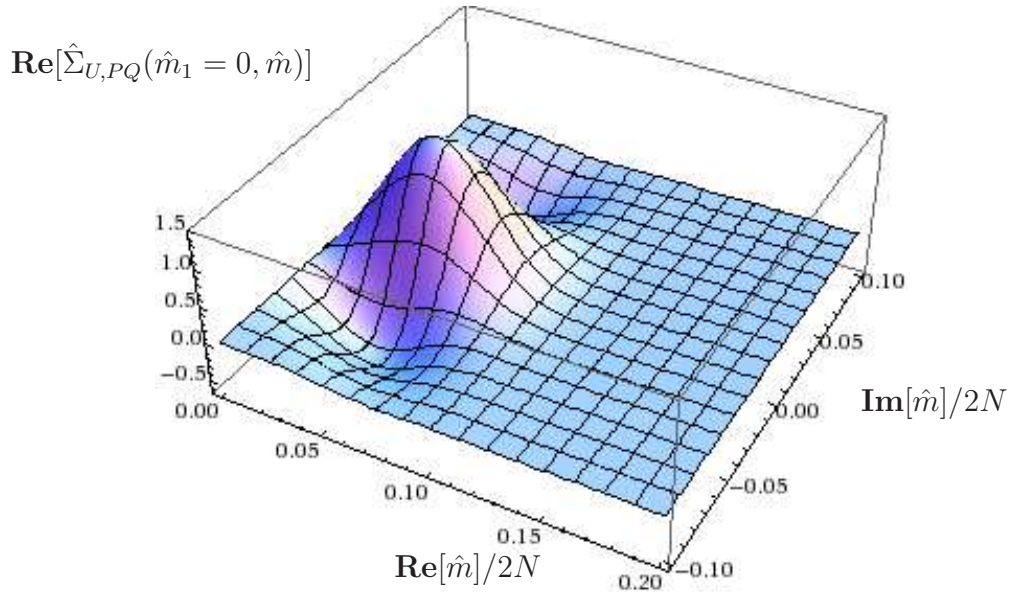,clip=,width=10cm}}
  \put(-1.7,-5.7){\bf\large Re$[\hm]/2N$}
  \put(4.5,-3.5){\bf\large Im$[\hm]/2N$}
  \put(-6.8,1.0){\bf \large Re$[\hat\Sigma_{U,PQ}(\hm_1=0,\hm)]$}
  \end{picture}
  \vspace{6cm}
  \end{center}
\caption{\label{fig:SigmaPQ-1} The real part of the unquenched contribution
  to the partially quenched
  condensate for one massless flavor, $\nu=0$ and $\hmu^2=3.84$ as obtained
  form the strong non-Hermiticity approximation. As in Figure
  2 the $x$ and $y$ axis are scaled by $2N=96$. Note that the result for
  Im$[\hat m]=0$ is the same as the full line in Fig. \ref{valmass}.}
\end{figure}

\begin{figure}[t]
  \unitlength1.0cm
  \begin{center}
  \begin{picture}(3.0,2.0)
  \put(-5.,-6.){
  \psfig{file= 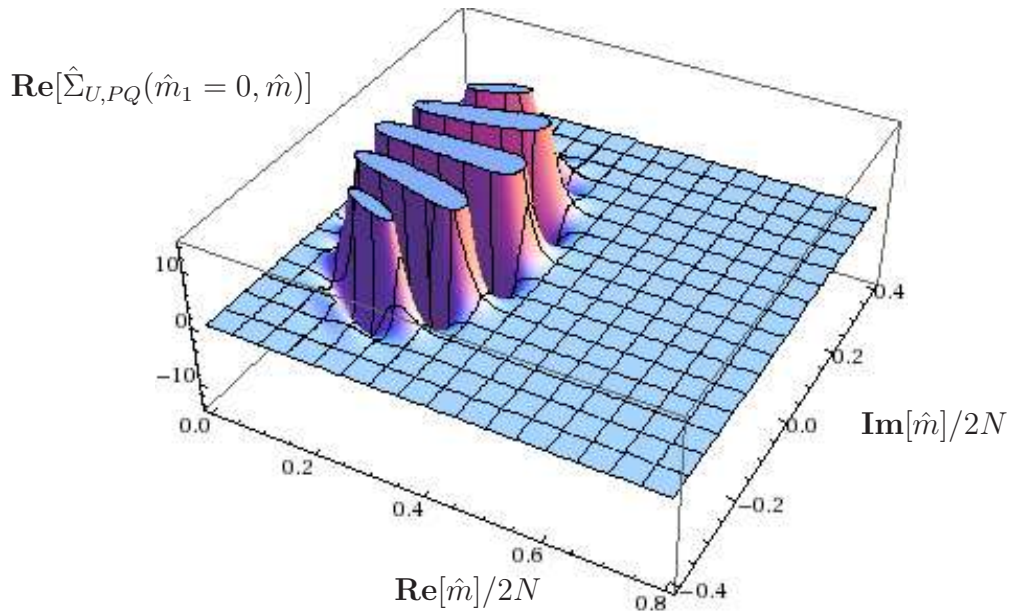,clip=,width=10cm}}
  \put(-1.7,-5.7){\bf\large Re$[\hm]/2N$}
  \put(4.5,-3.5){\bf\large Im$[\hm]/2N$}
  \put(-6.8,1.0){\bf \large Re$[\hat\Sigma_{U,PQ}(\hm_1=0,\hm)]$}
  \end{picture}
  \vspace{6cm}
  \end{center}
\caption{\label{fig:SigmaPQ-2} To illustrate how the oscillations of the
  spectral density develop we show (as above) the real part of the
  unquenched contribution to the partially
  quenched condensate for one massless flavor, $\nu=0$, but now
  $\hmu^2=4\times 3.84=15.36$. The maximal amplitude is about 50 times larger
  than the scale included in this plot.}
\end{figure}

For complex valence masses we obtain the valence quark mass dependence for $\nu =0$
and $m_1 =0$
\be
\hat\Sigma_{PQ}(\hat{m}_1=0,\hat m) 
& = & \hat\Sigma_Q(\hm)-\hat\Sigma_{U,PQ}(\hat{m}_1=0,\hat m)  \\
& = &
2u\hat m^* e^{-u \hat m^2/2}I_0(\hat m)K_1(u \hat m \hat m^*)-2/\hat m \nn \\
&&
+u^2 \hat m \hat m^* \hat m^*
\left(I_0(u \hat m \hat m^*)K_0(u \hat m \hat m^*)
    +I_1(u \hat m \hat m^*)K_1(u \hat m \hat m^*)  \right) +u\hat m. \nn
\label{SigmaUPQmms}
\end{eqnarray}
Two three dimensional plots of this result are shown in Figures
\ref{fig:SigmaPQ-1} and \ref{fig:SigmaPQ-2} for $\hmu^2 = 3.84$ and $\hmu^2 =
4\times 3.84 = 15.36$,  
respectively. Note that the $x$ and $y$ axis in the second plot have been
scaled up by the same factor of 4 as the chemical potential in order that 
the oscillating area covers approximately the same part of the two figures. 
We observe that the amplitude of the oscillations increase strongly with
increasing values of $\hmu$. 

\section{Conclusions}

It has been shown that the exact microscopic expression
for the dependence of the  chiral condensate on the dynamical quark mass
can be obtained by integrating 
the unquenched eigenvalue density over the complex plane. While the 
eigenvalue density depends strongly on the chemical potential, the 
chiral condensate does not depend on the chemical potential 
in the microscopic limit. 
Within a random matrix 
framework the cancellations leading to this result have
been connected to the orthogonality properties of orthogonal polynomials
in the complex plane and the factorization of the $\mu$-dependence of the 
polynomials in the microscopic limit. 
The simplifications occur even for finite size random matrices and 
the derivation is simpler in this case than in the  microscopic limit.
The proof of the microscopic limit was obtained by taking the microscopic
limit of relations derived for finite order polynomials. For completeness,
we also have given a derivation of the mass dependence of the chiral condensate
using complex contour integrations.

A second effort in this paper was to get a firmer grip on the properties of the
spectral density that lead to the $\mu$-independence of the 
chiral condensate.  This was achieved in the strong
non-Hermiticity limit for arbitrary $\nu$, where it was found
that  an integrability property of  the spectral density is responsible for
the desired cancellations. The interpretation of this property is that
the spectral density has been written as a derivative of the (complex) mass
dependence of the partially quenched chiral condensate. We have compared the
result for the partially quenched chiral condensate with earlier work, and
with exact results obtained by means of complex orthogonal polynomials.

A third result of this paper is the proof of a conjectured identity that
was used in the analysis of the partition function for one boson. In addition,
we have extended this result to arbitrary topological charge.

Finally, we wish to stress that  discontinuity
of the chiral condensate in the thermodynamic limit
is due to oscillating terms in the eigenvalue density
rather than an accumulation of eigenvalues on the imaginary axis: the 
Banks-Casher formula is not valid for QCD at nonzero chemical potential.
The nature of the oscillations exemplifies the difficulties encountered by
lattice QCD simulations at nonzero chemical potential where strong 
oscillations result in an exponentially suppressed phase factor
\cite{signprd} for $\mu>m_\pi/2$.

\noindent
{\bf Acknowledgments:} 
We wish to thank Poul Henrik Damgaard for useful discussions. This work was
supported in part by U.S. DOE Grant No. DE-FG-88ER40388 and by
the Carslberg Foundation (KS).    

\renewcommand{\thesection}{Appendix \Alph{section}}
\setcounter{section}{0}

\section{Direct derivation of the Microscopic Result}
\label{sec:directmicro}
 
In section \ref{sec:OSV} we have manipulated the non-universal finite $N$
expressions into a form which is suitable to take
the microscopic limit.  In this appendix we
show that it is also possible to take the microscopic limit from the start and
deal with the universal expressions throughout the argument.

The continuum limit of the sum over the index of the orthogonal polynomials
is given by
\be
\sum_{k=0}^{N-1}F(k) \to N\int_0^12t dt F(Nt^2).
\ee
The microscopic limit of the orthogonality relation reads
\be
\int_\mathbb{C} d^2 \hat z w(\hat z,\hat z^*;\hat\mu)
\hat{z}^{*\,-\nu}I_\nu(s \hat z^*) \hat{z}^{-\nu}I_\nu(t \hat z) =  
4\pi \hat \mu^2 s^{-1}
e^{2\hat \mu^2 t^2} 
\delta(s-t), 
\label{oi}
\ee
where we used the microscopic limit of the orthogonal polynomials given
in (\ref{p-mikro}).
The normalization can be verified by means of the large argument
asymptotic expansion of the modified Bessel functions.

The microscopic limit of the identity (\ref{J1}) is given by
\be 
 \int_\mathbb{C} d^2 \hat z\frac{w(\hat z,\hat z^*;\hat\mu) 
\hat z^{*\,-\nu}I_\nu(\hat z^* t)[\hat z^{-\nu}I_\nu(\hat z t)-
\hat m^{-\nu}I_\nu(\hat m t)]}{\hat z^2-\hat m^2} =0,
\label{Jo2}
\ee
and the microscopic limit of the expansion (\ref{kernelexp}) is given by
\be
\label{jmic}
\frac{\hat z^{-\nu}\hat I_\nu(\hat z)-\hat m^{-\nu}I_\nu(\hat m)}{\hat z^2-\hat
  m^2}=  
\int_0^1 A(\hat m,s)\hat z^{-\nu} I_\nu(\hat zs)s ds,
\label{expanmic}
\ee
with expansion coefficients equal to
\be
A(\hat m,s) =K_\nu(\hat ms)I_\nu(\hat m)-K_\nu(\hat m)I_\nu(\hat ms).
\ee 
Using the definition of a derivative we obtain
\be
{\hat m}^{-\nu}[I_\nu'(\hat m) -\frac \nu {\hat m}I_\nu(\hat m)] = 2 \hat m 
\int_0^1 A(\hat m,s)\hat m^{-\nu} I_\nu(\hat ms) ds.
\label{expander}
\ee

The microscopic spectral density for $N_f=1$ is given by \cite{O,AOSV}
\be
\rho_{N_f=1}^{(\nu)}(\hat{z},\hat{z}^*,\hat{m};\hat\mu)= \frac {w(\hat z,\hat z^*;\hat \mu)}{2\pi\hat \mu^2} 
\int_0^1 dt te^{-2\hat \mu^2 t^2} \left [\frac{
I_\nu(\hat z^*t) I_\nu(\hat zt)}{|\hat z|^{2\nu}} - 
\frac{I_\nu(\hat z^*t)I_\nu(\hat mt)}{|\hat z|^{2\nu}} 
\frac {I_\nu(\hat z)}{I_\nu(\hat m)}\right ].
\label{rhoNf1-micro}
\ee
By adding and subtracting $I_\nu(\hat{m}t)$ to $I_\nu(\hat{z}t)$ and $I_\nu(\hat{m})$ to
$I_\nu(\hat{z})$ the spectral density can be rewritten as
\be
\rho_{N_f=1}^{(\nu)}(\hat{z},\hat{z}^*,\hat{m};\hat\mu) & = & \frac {w(\hat z,\hat z^*;\hat \mu)}{2\pi\hat \mu^2} 
\int_0^1 dt te^{-2\hat \mu^2 t^2}  
\frac{I_\nu(\hat z^*t)}{\hat z^{*\,\nu}} \nn \\ 
 && \times 
\left ( 
[\hat{z}^{-\nu}I_\nu(\hat zt)
-\hat m^{-\nu} I_\nu(\hat m t)] 
- [\hat{z}^{-\nu}I_\nu(\hat z)-\hat m^{-\nu}I_\nu(\hat m)]
\frac{I_\nu(\hat mt)} {I_\nu(\hat m)}\right ).
\label{eq48}
\ee

The chiral condensate is given by
\be
\hat \Sigma(\hat m) = \int_\mathbb{C} d^2 \hat z 
\frac {\hat m}{\hat m^2-\hat z^2}
\rho_{N_f=1}^{(\nu)}(\hat{z},\hat{z}^*,\hat{m};\hat\mu).
\ee
Because of the expansion (\ref{expanmic}) 
and the orthogonality relation (\ref{oi}),
the first bracketed term in Eq. (\ref{eq48})
does not contribute to the chiral condensate resulting in
\be
\hat\Sigma(\hat m) = \hat m\int_\mathbb{C} d^2 \hat z \frac{w(\hat  z,\hat z^*;\mu)}{2\pi \hat \mu^2}
\int_0^1 dt t e^{-2\hat \mu^2 t^2} 
\frac {\hat z^{-\nu}I_\nu(\hat z) - \hat m^{-\nu}I_\nu(\hat m)}{\hat z^2-\hat m^2}
\frac{I_\nu(\hat z^*t)}{\hat z^{*\, \nu}} \frac{I_\nu(\hat mt) }{I_\nu(\hat m)}.
\ee 
Finally, after inserting the expansion (\ref{expanmic}) and
applying the orthogonality relation 
(\ref{oi}), the derivative relation (\ref{expander}) results in
\be
\hat\Sigma(\hat m) = \frac {I_\nu'(\hat m)}{I_\nu(\hat m)}-\frac \nu {\hat{m}},
\ee
which is the correct $\mu$-independent chiral condensate
after including the contribution from the zero modes.


\section{Chiral condensate from complex contour integrations}
\label{sec:contour}

In this appendix we derive the chiral condensate by integration
over the eigenvalue density using complex contour integrations. 
We only work out the case of $\nu = 0$.  

Let us start by writing the chiral condensate for $N_f=1$ as 
\be
\hat\Sigma(\hat m) = \int_0^1 dt \hat\Sigma(\hat m,t)
\ee
where
\be
\hat\Sigma(\hat m,t) = \int_\mathbb{C} d^2 \hat z \frac 1{\hat z+\hat m} 
\tilde\rho_{N_f=1}^{(0)}(\hat z,{\hat z}^*,t; \hm, \hmu),
\ee
and 
\be
\tilde\rho_{N_f=1}^{(0)}(\hat z,{\hat z}^*,t;\hat  m, \hat \mu) = \frac{\hat z{\hat z}^*}{2\pi\hat  \mu^2}
e^{-(\hat z^2+{\hat z}^{* 2})/8\hat \mu^2} K_0(|\hat z|^2/4\hat \mu^2) 
te^{-2\hat \mu^2 t^2} I_0({\hat z}^* t) I_0(\hat mt)\frac{I_0(\hat m)- I_0(\hat z)}{I_0(\hat m)},
\ee
which follows from (\ref{rhoNf1-micro}) after use of (\ref{Jo2}). Since 
the chiral condensate is real,
there is no need to consider imaginary
contributions to $\hat\Sigma(\hat m,t)$.

We write $\hat z= x+i y$ and do the integral over $y$ by a contour integration.
Since we perform the integration over $y$ by a contour integration, we have to
distinguish the cases $x< -\hat m$ and $ x > -\hat m$ 
and decompose $I_0(\hat z)$ as 
\be
I_0(\hat z) = \frac 1\pi \frac {\sqrt{\hat z}}{\sqrt{-\hat z}}
( K_0(\hat z) - K_0(-\hat z)).
\ee
For $I_0(\hat z^*t)$, a similar decomposition is used.
The pole is at $y=i(x +\hat m)$ so that
\be
\hat z \to -\hat m\qquad {\rm and} \qquad \hat z^* \to 2x+\hat m.
\ee
Completing the contour integral we obtain for the pole contribution
\be
\hat\Sigma(\hat m,t) &=& \frac {-\hat mt}{\pi\hat \mu^2}\int dx
(2x+\hat m)e^{-\frac{2x^2+\hat m(2x+\hat m)}{4\hat \mu^2}-2\hat \mu^2t^2}
K_0(-\frac{\hat m(2x+\hat m)}{4\hat \mu^2})\frac{I_0(\hat mt)}{I_0(\hat
  m)}\nn \\ 
&&\times \left [ \theta(x+\hat m)[-i\sgn(2x+\hat m)K_0((2x+\hat m)t)I_0(\hat m) +iK_0(\hat m)I_0((2x+\hat m)t)]
\right . \nn  \\ && \left .
-\theta(-x-\hat m)[i\sgn(2x+\hat m)K_0(-(2x+\hat m)t) I_0(\hat m) - iK_0(-\hat
m)I_0((2x+\hat m)t)]\right ]. 
\ee
There is also a contribution from the jump across the cuts along the 
imaginary axis. However, since these contributions are purely imaginary
we do not further analyze them.

To disentangle the different contributions, we rewrite the hyperbolic Bessel 
functions as 
\be
K_0(x) = K_0(|x|) - \pi  i I_0(|x|).
\ee
After the cancellation of two terms for $x+\hat m<0$ we observe that the 
integrand is given by the same expression as for $x+\hat m >0$.
Using $y=2x +\hat m$ as new integration variable the expression for
${\rm Re}(\hat\Sigma(\hat m,t))$ simplifies to
upon integration,
\be
{\rm Re}(\hat\Sigma(\hat m,t)) &=& 
\frac {\hat mt}
{2\hat \mu^2}\frac{I_0(\hat mt)}{I_0(\hat m)}e^{-2\hat \mu^2 t^2-\frac {\hat m^2}{8\hat \mu^2}}
\int dy ye^{-\frac{y^2}{8\hat \mu^2}} \nn\\ &&\times
\left [
 \theta(y) I_0\left ( \frac{\hat m|y|}{4\hat \mu^2}\right)(I_0(\hat m) \sgn(y) K_0(|y|t)
- K_0(\hat m)  I_0(|y|t)) \right.
\nn\\ && \left. \ \ \   + \theta(-y) K_0\left ( \frac{\hat m|y|}{4\hat \mu^2}\right)I_0(\hat m) \sgn(y) I_0(|y|t)
\right ] .
\ee
The integrals can be simply rewritten as
\be
{\rm Re}(\hat\Sigma(\hat m,t)) &=& 
\frac {\hat mt}{2\hat \mu^2}\frac{I_0(\hat mt)}{I_0(\hat m)}
e^{-2\hat \mu^2 t^2-\frac {\hat m^2}{8\hat \mu^2}}
\int_0^\infty dy ye^{-\frac{y^2}{8\hat \mu^2}}  \\ && \times 
\left [
[ I_0\left ( \frac{\hat my}{4\hat \mu^2}\right)K_0(yt)
+ K_0\left ( \frac{\hat my}{4\hat \mu^2}\right)  I_0(yt)]I_0(\hat m)
- I_0\left ( \frac{\hat my}{4\hat \mu^2}\right) I_0(yt)K_0(\hat m)
\right ] .\nn
\ee
The integral of the terms multiplying $I_0(\hat m)$ was evaluated
in \cite{signprd} whereas the integral containing the last term in the 
above equation can be found in \cite{Gradshtein}.
We finally find
\be
{\rm Re}(\hat\Sigma(\hat m,t)) &=& 
{2\hat mt}\frac{I_0(\hat mt)}{I_0(\hat m)}[K_0(\hat mt)I_0(\hat m) - K_0(\hat m) I_0(\hat mt)]
\ee
in agreement with results derived in subsection \ref{subsec:Sigmak}.

\section{Spectral Density as a Total Derivative}
\label{sec:appB}

In this appendix we obtain the chiral condensate by writing the spectral
density as a total derivative. This will be worked out in the 
strong non-hermiticity limit only, first for $\nu = \frac 12$ and then
for general $\nu$.

\subsection{Strong Non-Hermiticity}

The microscopic limit of the spectral density (\ref{rhoNf1}) for $N_f=1$ is
\cite{O,AOSV}
\be\label{rhoNf1mikro}
\rho_{N_f=1}^{(\nu)}(\hat{z},\hat{z}^*,\hat{m};\hat\mu) 
& = & \frac{|\hat{z}|^2}{2\pi \hat\mu^2} 
K_\nu\left(\frac{|\hat{z}|^2}{4\hat\mu^2}\right)
\mbox{e}^{-\frac{\hat{z}^2+\hat{z}^{*\,2}}{8\hat\mu^2}}\\
&& \times\left(\int_0^1 dt \ t \ \mbox{e}^{-2\hat\mu^2 t^2}
I_\nu(\hat{z} t)I_\nu(\hat{z}^* t)-\frac{I_\nu(\hat{z})}{I_\nu(\hat{m})}\int_0^1 dt \ t \ \mbox{e}^{-2\hat\mu^2 t^2}I_\nu(\hat{m} t)I_\nu(\hat{z}^* t)\right)\nn
\ee
The term with the first integral is the quenched density and 
the second term gives
the effect of unquenching, which motivates the notation 
\be
\rho_{N_f=1}^{(\nu)} = \rho_Q^{(\nu)} - \rho_U^{(\nu)}.
\label{qudec}
\ee
It is easily checked that $\rho_{N_f=1}$ vanishes at $\hat{z}=\hat{m}$ since $\rho_Q =
\lim_{\hat{m}\to\hat{z}}\rho_U$. This trivial observation which holds since the fermion
determinant vanishes when an eigenvalue is equal to the quark mass will be
very useful below.  

In the limit of strong non-Hermiticity $\hat\mu^2\gg 1$ with
$\hat{x}/(2\hat\mu^2)<1$ and $(\hat{x}+\hat{m})/(4\hat\mu^2)<1$ such that the saddle 
points of the $t$-integrations are inside the interval $[0,1]$, 
the integrals
in (\ref{rhoNf1mikro}) can be approximated by  
\be
\int_0^1 dt \, t e^{-2\hat\mu^2 t^2} I_\nu((\hat{x}-i\hat{y})t)I_\nu(\hat{m}t)  
&\approx &
 \int_0^\infty dt \, t  e^{-2\hat\mu^2 t^2} I_\nu((\hat{x}-i\hat{y})t) 
 I_0(\hat{m}t)\nn \\
 & = & \frac 1{4\hat\mu^2}e^{\frac{(\hat{x}-i\hat{y})^2+\hat{m}^2}{8\hat\mu^2}}  
I_\nu\left(\frac{\hat{m}(\hat{x}-i\hat{y})}{4\hat\mu^2}\right).
\label{largemu-app}
\ee
Within this limit the quenched spectral density for topological charge $\nu$
reduces to 
\be
\rho_Q^{(\nu)} = \frac {2u^2}{\pi} \hat z \hat z^* 
K_\nu(u\hat  z\hat  z^*) I_\nu(u\hat z\hat  z^*),
\ee
while the unquenched part reads
\be
\rho_U^{(\nu)} =\frac {2u^2}{\pi}\hat z\hat z^*K_\nu(u\hat z\hat  z^*) I_\nu(u\hat m\hat  z^*)\frac {I_\nu(\hat z)}{I_\nu(\hat m)}
e^{u(\hat m^2-\hat z^2)/2}
\ee
where we introduced the abbreviation $u\equiv 1/4\hat \mu^2$. Below we will
express $\rho$ as a total derivative and show that the resulting chiral
condensate is independent of $\hat\mu$. Because the chiral condensate is
independent of 
$\hat\mu$ we recover the full microscopic chiral condensate even though we 
work in the limit of strong non-Hermiticity. Before we present the argument for 
general topological index we first work through the simpler case $\nu=1/2$.

\subsubsection{The case of $\nu = \frac 12$}

The microscopic spectral density for $\nu = \frac 12$ and $N_f = 1$ is given
by 
\be
\rho_{N_f=1}^{(\nu= \frac 12) }(\hat z,\hat z^*,\hat m;\hat \mu)
 = \rho^{(\nu= \frac 12)}_Q(\hat z,\hat z^*;\hat \mu) 
-\rho^{(\nu= \frac 12)}_U(\hat z,\hat z^*,\hat m;\hat \mu) 
\ee
with (using the notation $u \equiv 1/4\hat \mu^2$)
\be
\rho_{\rm U}^{(\frac 12)} = \frac u{\pi } e^{-\frac u2\hat  z^2
-u{\hat z{\hat z}^*} +\frac u2 \hat{m}^2 }
\frac{e^{\hat z}-e^{-\hat z}}{e^{\hat m} - e^{-\hat m}} (e^{u{\hat m{\hat z}^*}} - 
e^{-{u\hat  m\hat  z^*}}).
\ee
and 
\be
\rho_{\rm Q}^{(\frac 12)} = \lim_{\hat m\to\hat  z} \rho_U =\frac u{\pi} 
(1-e^{-2 u{\hat z\hat  z^*}}).
\ee
It is straightforward to integrate $\rho_U$ with respect to $\hat z^*$:
\be 
\rho_U^{(\frac 12)} &=& \frac 1\pi \del_{\hat z^*}[\frac{f(\hat z,\hat m) e^{
u\hat  z^*(\hat m-\hat z)} -1}{\hat m-\hat z}]+
             \frac 1\pi \del_{\hat z^*}[\frac{f(\hat z,\hat m) e^{-u\hat
                 z^*(\hat m+\hat z)} +1}{\hat z+\hat m}] 
\label{rhoUasdz*}
\ee
with 
\be
f(\hat z,\hat m) = \frac{e^{\hat z}-e^{-\hat z}}{e^{\hat m}-e^{-\hat m}}e^{\frac u2(\hat m^2-\hat z^2)}.
\ee
The limit $\hat m \to \hat z$ 
and the derivative with respect to $\hat z^*$ can be interchanged
for smooth functions. However, taking the limit first can generate 
singularities that result in spurious contributions after
differentiating with respect to $\hat z^*$. Taking the limit
$\hat{m}\to\hat{z}$ in (\ref{rhoUasdz*}) by naive differentiation leads to 
\be
\rho_Q^{(\frac 12)}= \frac 1\pi \del_{\hat z^*} \left [ 
u\hat z +u\hat z^* - \frac{\cosh \hat z}{\sinh\hat  z}+ \frac {e^{-2u\hat z\hat  z^*}+1}{2\hat z}
\right ] .
\ee
To eliminate the spurious singular terms  we subtract
\be
\tilde \rho_Q^{(\frac 12)} = -\frac 1\pi \del_{\hat z^*}[\frac{\cosh\hat  z}{\sinh\hat  z} -\frac 1{\hat z}]
\ee
from the quenched spectral density. The chiral condensate can then be
expressed as
\be
\hat\Sigma(\hat m) &=& \int d\hat x d\hat y \frac 1{\hat m+\hat z}(\rho_Q^{(\frac 12)} 
- \tilde \rho_Q^{(\frac 12)} -\rho_U^{(\frac 12)}). 
\ee
The contribution of the unsubtracted derivatives vanishes at $\hat z= \hat m$. After 
partial integration we thus find
\be
\hat\Sigma(\hat m) = \frac{\cosh \hat m}{\sinh\hat  m} -\frac 1{\hat m}=
\frac{I_{3/2}(\hat m)}{I_{1/2}(\hat m)}. 
\ee
After adding the contribution from the zero modes given by $\nu /\hat m$ we
find 
\be
\hat\Sigma(\hat m) &=&\frac {I_{3/2}(\hat m)}{I_{1/2}(\hat m)} + \frac 1{2\hat m}\nn\\
&=&\frac {I_{1/2}'(\hat m)}{I_{1/2}(\hat m)}
\ee
in agreement with the result for arbitrary topological charge to 
be discussed next.

\subsubsection{The general $\nu$ case}

We now turn to the case with arbitrary $\nu$. Similar to the case $\nu =
\frac 12$ above, we express 
$\rho_U^{(\nu)}$ as 
\be
\rho_U^{(\nu)} = \frac 1\pi \del_{z^*} \frac{F(\hat z,\hat z^*,\hat m)}{\hat m^2-\hat
  z^2}, 
\label{gensigu}
\ee
with
\be
F(\hat z,\hat z^*,\hat m)&=& 2 \hat z e^{u(\hat m^2-\hat z^2)/2} \frac {I_\nu(\hat z)}{I_\nu(\hat m)}
   [\hat m\hat  z^* u K_\nu(u\hat z\hat  z^*)I_\nu'(u\hat m\hat  z^*)-\hat z
   \hat z^*uI_\nu(u\hat m\hat  z^*)K_\nu'(u\hat z\hat  z^*)]
-2\hat z,\nn \\
&\equiv& F_A(\hat z,\hat m) G(\hat m\hat  z^*,\hat z\hat  z^*) -2\hat z.
\label{genF}
\ee 
Here, 
\be
F_A(\hat z, \hat m) = 2 \hat z \exp(u(\hat m^2-\hat z^2)/2) I_\nu(\hat z)/I_\nu(\hat m),
\ee
and
\be 
G(\hat m\hat  z^*,\hat z\hat  z^*) =\hat m\hat  z^* u K_\nu(u\hat z\hat  z^*)
I_\nu'(u\hat m\hat  z^*)-\hat z  \hat z^*uI_\nu(u\hat m\hat  z^*)
K_\nu'(u\hat z\hat  z^*).
\ee
The term $2\hat z$ has been subtracted to eliminate singularities at
$\hat z=\pm \hat m$.
Using the Wronskian identity
\be
I_\nu'(x) K_\nu(x) - I_\nu(x) K_\nu'(x) = \frac 1{x}
\ee
one easily shows that
\be
\left . G(\hat m\hat  z^*,\hat  z\hat  z^*)F_A(\hat z, \hat m) 
\right|_{\hat m=\pm\hat  z} = 2\hat z.
\ee
Therefore,
\be
F_A(\hat z,\hat m) G(\hat m\hat  z^*,\hat z\hat  z^*) -2\hat z 
= (\hat m^2-\hat z^2)h(\hat m,\hat z,\hat z^*)
\label{reason}
\ee
so that
\be
\del_{\hat m}[F_A G -2\hat z]_{\hat z=\hat z^*=\hat m} = -\del_{\hat z}
[F_A G -2\hat z]_{\hat z=\hat z^*=\hat m}
\label{consist}
\ee
with $h(\hat m, \hat z,\hat z^*)$ a function that is regular at $\hat m = \hat z$.

The quenched part of the spectral density is given by
\be
\rho_Q^{(\nu)}(\hat z,\hat z^*;\hat\mu) = \lim_{\hat m\to \hat z} \rho_U^{(\nu)}(\hat z,\hat z^*,\hat
m;\hat\mu).
\ee
Naively interchanging this limit with $\del_{\hat z^*}$ results in
\be
\rho_Q^{(\nu)}(\hat z,\hat z^*) = \frac 1\pi \left (
\del_{\hat z^*}\frac {\left . F_A(\hat z,\hat z)\del_{\hat m} 
G(\hat m \hat z, \hat z \hat z^*)\right|_{\hat m=\hat z}}{2\hat z}+
\del_{\hat z^*}\frac {\left .G(\hat m \hat z, \hat z \hat z^*)
\del_{\hat m} F_A(\hat z, \hat m)\right|_{\hat m=\hat z}}{2\hat z} \right ).
\ee
However, the second term may give rise to singular contributions that
would be absent if we would have differentiated with respect
to $\hat z^*$ before taking the limit. Therefore these contributions have
to be subtracted so that the quenched spectral density is given by
\be 
\rho_Q^{(\nu)}(\hat z,\hat z^*) = \frac 1\pi \left (
\del_{\hat z^*}\frac {\del_{\hat m}[ F_A(\hat z,\hat m) 
G(\hat m \hat z, \hat z \hat z^*)]_{\hat m=\hat z}}{2\hat z}-
\del_{\hat z^*}{\rm Sing}\left [\frac {\left . 
G(\hat m \hat z, \hat z \hat z^*)\del_{\hat m} F_A(\hat z,\hat m)
\right|_{\hat m=\hat z}}{2 \hat z}\right ] \right ).
\label{rhoqsub}
\ee
The chiral condensate can be written as
\be
\hat\Sigma(\hat m) &=& \int d^2 \hat z \frac 1{\hat m+\hat z}\frac 1\pi 
\del_{\hat z^*}
\left [  \frac {\del_{\hat m}[ F_A(\hat z,\hat m)
G(\hat m \hat z, \hat z \hat z^*))]_{\hat m=\hat z}}{2\hat z}
\right . \\ && \left .
-{\rm Sing}\left [\frac {\left . G(\hat m \hat z, \hat z \hat z^*)
\del_{\hat m} F_A(\hat z,\hat m)
\right|_{\hat m=\hat z}}{2\hat z}\right ]
- \frac { F_A(\hat z,\hat z)G(\hat m \hat z, \hat z \hat z^*)-2 \hat z}
{\hat m^2-\hat z^2}
\right] .\nn
\ee
After partial integration with respect to $z^*$ we find
\be
\hat\Sigma(\hat m) &=& 
-\left [  \frac { \del_{\hat m} [ 
F_A(\hat z,\hat z) G(\hat m \hat z, \hat z \hat z^*) ]_{\hat m=\hat z}}{2\hat z}\right
. \\ && \left . 
-{\rm Sing}\left [\frac {\left . G(\hat m \hat z, \hat z \hat z^*)
\del_{\hat m} F_A(\hat z, \hat m)
\right|_{\hat m=\hat  z}}{2 \hat z}\right ]
+ \frac {\del_{\hat z}[F_A(\hat z, \hat z)
G(\hat m \hat z, \hat z \hat z^*) -2\hat z]}{2 \hat m} 
\right ]_{\hat z=\hat z^*=-\hat m}.\nn
\ee
After using (\ref{consist}) only the singular terms remain. 
Because of (\ref{reason}) there is 
no singularity at $\hat z = 0$ 
after using (\ref{consist}) 
\be 
\hat\Sigma(\hat m) =  
 {\rm Sing}\left [
\frac {G(\hat m \hat z,\hat z\hat z^*) 
\del_{\hat m}  \left .
F_A(\hat z, \hat m)\right |_{\hat m = \hat z}
}{2 \hat z}
\right ]_{\hat  z=\hat z^*=-\hat m}
= \frac {I'_\nu(\hat m)}{I_\nu(\hat m)} -\frac \nu {\hat m},
\ee
which gives the correct result after  
including the contribution from the zero modes.

\end{document}